\documentclass[default,iicol]{sn-jnl}

\usepackage{graphicx}%
\graphicspath{ {figures/} }
\usepackage{subcaption}
\usepackage{multirow}%
\usepackage{amsmath,amssymb,amsfonts}%
\usepackage{amsthm}%
\usepackage{mathrsfs}%
\usepackage{placeins}
\usepackage[title]{appendix}%
\usepackage{xcolor}%
\usepackage{textcomp}%
\usepackage{manyfoot}%
\usepackage{booktabs}%
\usepackage{algorithm}%
\usepackage{algorithmicx}%
\usepackage{algpseudocode}%
\usepackage{listings}%
\usepackage{mathtools}
\usepackage{dutchcal}

\newcommand{\letterref}[2]{\hyperref[#1]{\autoref*{#1}\textbf{#2}}}

\theoremstyle{thmstyleone}%
\theoremstyle{thmstyletwo}%

\theoremstyle{thmstylethree}%

\begin{document}

\title[Mitigating Biases in Collective Decision-Making: Enhancing Performance in the Face of Fake News]{Mitigating Biases in Collective Decision-Making: Enhancing Performance in the Face of Fake News}

\author*[1,2,3]{Axel Abels}
\email{axel.abels@ulb.be}
\author[1,2,3]{Elias Fernandez Domingos}
\email{elias.fernandez.domingos@ulb.be}
\author[2,3]{Ann Now{\'e}}
\email{ann.nowe@vub.be}
\author*[1,2,3,4]{Tom Lenaerts}
\email{tom.lenaerts@ulb.be}

\affil[1]{\orgdiv{Machine Learning Group}, \orgname{Universit{\'e} Libre de Bruxelles},
            \orgaddress{Boulevard du Triomphe, CP 212},
            \city{Brussels}, \postcode{1050}, \country{Belgium}}
\affil[2]{\orgdiv{AI Lab}, \orgname{Vrije Universiteit Brussel},
            \orgaddress{Pleinlaan 2}, \city{Brussels}, \postcode{1050}, \country{Belgium}}
\affil[3]{\orgdiv{FARI Institute}, \orgname{Universit{\'e} Libre de Bruxelles - Vrije Universiteit Brussel}, \city{Brussels},
            \postcode{1000},
            \country{Belgium}}
\affil[4]{\orgdiv{Center for Human-Compatible AI}, \orgname{UC Berkeley},
            \orgaddress{2121 Berkeley Way},
            \postcode{94720 Berkeley, CA},
            \country{USA}}





\abstract{
Individual and social biases undermine the effectiveness of human advisers by inducing judgment errors which can disadvantage protected groups. In this paper, we study the influence these biases can have in the pervasive problem of fake news by evaluating human participants' capacity to identify false headlines. By focusing on headlines involving sensitive characteristics, we gather a comprehensive dataset to explore how human responses are shaped by their biases. Our analysis reveals recurring individual biases and their permeation into collective decisions. We show that demographic factors, headline categories, and the manner in which information is presented significantly influence errors in human judgment. We then use our collected data as a benchmark problem on which we evaluate the efficacy of adaptive aggregation algorithms. In addition to their improved accuracy, our results highlight the interactions between the emergence of collective intelligence and the mitigation of participant biases. 
}


\maketitle

Cognitive biases are systematic errors in judgment and decision-making resulting from cognitive limitations, individual preferences, and/or inappropriate heuristics \citep{tversky1974judgment}.
 When human groups deliberate, individuals tend to transmit these biases to others, which may lead the group as a whole to make sub-optimal choices \citep{janis2008groupthink}. Such biases can pose a significant threat to accurate and fair decision-making, especially in situations that involve sensitive attributes. In settings wherein group decisions play a pivotal role --- from medical diagnostics to crowdsourcing --- an understanding of these biases, how they affect the integrity and efficiency of collective decisions, and how they can be countered is paramount. 

Existing literature underscores the profound influence of stereotypes, biases, and decision-making across multiple facets of life, from choices in recruitment \cite{rooth2010automatic} to judiciary decisions \cite{lin2020limits,dressel2018accuracy}. While tools and studies like the Stereotype Content Model \citep{fiske2002model,cuddy2009stereotype}, Implicit Association Tests \citep{greenwald1998measuring,greenwald2006implicit} offer crucial insights into measures of biases, there remains a considerable scope for further exploration. This includes the understanding of interactions between bias mitigation and the emergence of collective intelligence. 

\begin{figure*}[ht!]
\centering
 
 \includegraphics[width=.8\textwidth]{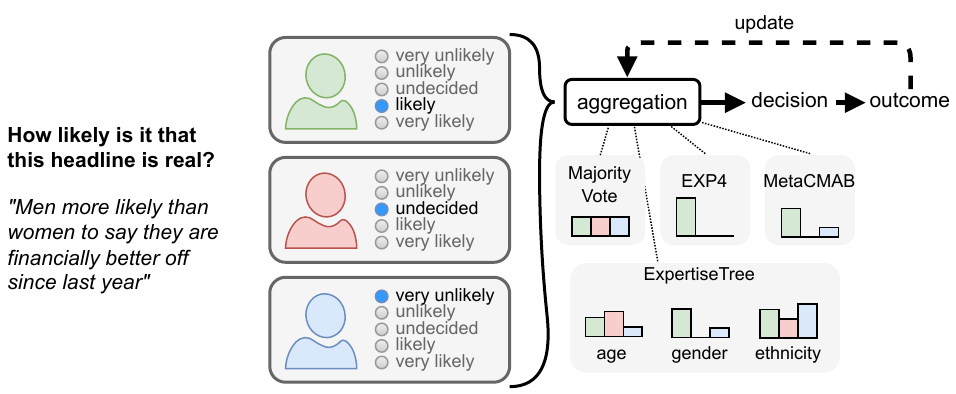}

 \caption{%
 Overview of the Collective Decision-Making problem studied in this work. 
  Participants (identified here by the colors green, red and blue) in the experiment were presented with a sequence of headlines and were asked to estimate the likelihood that they were true. These diverging opinions are then aggregated to reach a collective decision. The aggregation is iteratively optimized by comparing it to a ground truth. Different aggregations weigh opinions differently, see \autoref{sec:algorithms} for algorithmic details. Example weights for the three participants are given for each approach. A majority vote values all experts uniformly, resulting in even bars. EXP4 \cite{auer2002nonstochastic} selects a single participant (i.e., weight is concentrated on a single participant) whose opinion is followed. 
  MetaCMAB \cite{abels2023dealing} distributes weights more evenly, correlating them with performance to enhance the collective decision-making process. ExpertiseTree \cite{pmlr-v202-abels23a} similarly distributes weights, but will learn distinct models for each headline category if this proves beneficial.}\label{fig:problem-illustration}
\end{figure*}

Pooling human expertise through a centralized platform may mitigate biases (e.g., confidence bias \citep{harvey1997confidence}, biases with temporal effects \citep{baddeley1993recency,jones2001positive}, or illusory correlations \citep{pohl2004cognitive}) that affect the judgment of participants in a collective decision-making task. Works such as those of de Condorcet \cite{condorcet1785essay}, Rawles \citep{rawls1971atheory}, and Surowiecki \citep{surowiecki2005wisdom} have provided a theoretical understanding of how collective decisions can bring about improvements over individuals. Modern algorithms (e.g., \citep{luo2018efficient,abels2023dealing,pmlr-v202-abels23a}), while drawing inspiration from those seminal works, address the complexities of aggregating possibly dissenting opinions by dynamically adapting to varying levels expertise. While those works provide a theoretical analysis of the performance of these algorithms, empirical performance has so far been primarily established on synthetic experts \citep{pang2018dynamic,abels2020expert,abels2023dealing,pmlr-v202-abels23a}. It is therefore not obvious that the assumptions they rely on hold for human experts, and thus, how these methods perform in a real situation. 

The primary purpose of this study is to address this gap by experimentally evaluating the performance of both static (e.g., majority voting) and adaptive (EXP4 \citep{auer2002nonstochastic}, MetaCMAB \citep{abels2023dealing}, and ExpertiseTree \citep{pmlr-v202-abels23a}) aggregation algorithms in settings involving human advisers. Specifically, we evaluate their capacity to mitigate biases and enhance the efficacy of collective decision-making.

Our experimental design\footnote{This experiment has been preregistered \citep{osfreg}}, inspired by crowdsourced fact checking \cite{sethi2017crowdsourcing,wei2022combining,allen2021scaling}, involves human participants tasked with discerning the authenticity of news headlines, particularly those that involve sensitive groups (see \autoref{fig:problem-illustration}). In the experiment, participants encountered a sequence of headlines, half of which were authentic and half of which were altered by switching the sensitive groups involved. Without being informed of the proportion of authentic headlines, they were asked to rate the likelihood of each headline being genuine. By contrasting the true outcome with opinions provided by participants, adaptive algorithms can optimize the aggregation of human opinions, possibly conditioned on the problem category.

 This setup enabled us to not only identify potential bias patterns in participant responses but also to assess the effectiveness of various collective decision-making algorithms in real-world scenarios. Crucially, our study demonstrates the potential of state-of-the-art machine learning algorithms to tailor aggregations in a way that mitigates biases and enhances collective intelligence.

To summarize, our contributions are 
\begin{itemize}
    \item A dataset containing participant responses to a collection of news headlines involving sensitive characteristics, providing a rich source for analyzing human biases in decision-making.
    \item A comprehensive exploration of these responses, focusing on potential biases and the presence of stereotyping.
    \item An analysis of the performance of collective decision-making algorithms, with an emphasis on the interaction between bias mitigation and the emergence of collective intelligence.
\end{itemize}

\section{Overview}
We designed an online experiment (see Methods for more information) which emulates the use of human fact checkers as a tool to combat the spread of fake news on social media. Participants were presented with a sequence of headlines for which they were asked to rate the likelihood of the headline being real. In order to elicit bias, we focused on headlines which involved sensitive groups. 

In particular, we curated headlines from \citep{DVN/SYBGZL_2018,mazumder2014news,Medscape} based on the following criteria:
\begin{itemize}
 \item  The headline (implicitly or explicitly) contrasts two sensitive groups (e.g., \textit{``Men more likely than women to say they are financially better off since last year"})
 \item The headline should present a clear negative or positive outcome. For example, it is not clear whether \textit{``Poll: Kanye more popular with whites than nonwhites"} is positive or negative, but \textit{``African-Americans, Hispanics, dying at faster rate of fentanyl overdoses than whites: analysis"} is clearly a negative outcome for African-Americans and Hispanics.
\end{itemize}

Each headline was characterized by a sensitive group (thus whether it concerned a gender group, an ethnic group, or an age group), by a sentiment (positive or negative), and by a truth value (whether the headline was real or altered). For details, see Methods.

We sequentially presented one of $5$ distinct sets of headlines to each participant and asked them to assess the likelihood of each being real using a discrete score: For each headline, participants were asked ``How likely is it that
this headline is real?" and given $5$ choices (see \autoref{fig:problem-illustration}): \{``very unlikely", ``unlikely", ``undecided", ``likely", ``very likely"\}. 

Once collection of all participant responses was completed, these responses were used in simulated collective decision-making tasks (see \autoref{sec:CDM_process}). For each simulation, we sampled uniformly $N \in \{2,4,6,...,36\}$ participants who answered the same set of headlines. Their answers to each headline were iteratively fed into the studied aggregation algorithms. These algorithms then aggregated the responses from this subset to reach a collective decision. Following each decision, the algorithms were presented with the actual type of the headlines (i.e., genuine or altered). This feedback loop (see \autoref{fig:problem-illustration}) allows the algorithms to adapt and refine the decision-making processes to improve accuracy in subsequent iterations on novel headlines\footnote{Headlines are not repeated within a simulation. Any decision made is therefore the result of training on a history that does not include the headlines on which the decision-making quality is estimated. Specifically, at time $t$ we evaluate the quality of a decision made by a model trained on the data experienced at times $1$ to $t-1$. This allows us to test the effectiveness of models on data they have not encountered before, analogously to training and test sets which are more typical of a supervised learning setting. }. 

Crucially, the manner in which the various algorithms studied here assimilate experiences varies. For instance, the weighted majority vote (WMV) algorithm is static and does not adapt based on past outcomes. Considering the sensitive attributes involved in this setting, our intent is to counteract biases. This is achieved through:
\begin{enumerate}
    \item A collective approach: Rather than pinpointing the single best participant, we optimize a weighted average of expertise. Intuitively, using a weighted average —--instead of singling out one participant —-- may ensure that individual errors offset each other. 
    \item An active bias counteraction method: Recognizing that biases may cause varying expertise levels dependent on the sensitive group, we suggest learning aggregates tailored to the sensitive groups we encounter. This can be realized by implementing an ExpertiseTree \citep{pmlr-v202-abels23a}. Essentially, this decision tree's leaves contain an aggregator optimized for a subset of sensitive groups. For example, if headlines related to gender benefit from a distinct aggregate, the ExpertiseTree algorithm provides an adapted aggregator to handle that need.
\end{enumerate}

\noindent Our data analysis led to several significant findings:

\begin{itemize}
    \item Algorithms accounting for participant bias (like ExpertiseTree) consistently achieve collective intelligence, in doing so these algorithms significantly mitigate the biases present in individual participants.
    \item Participants tend to be less skeptical towards headlines involving age and more skeptical towards headlines involving ethnicity.
    \item When performance differences between groups occur, they tend to occur on headlines involving those groups. For example, men and women perform comparably on ethnic or age-related headlines, but differ on gender headlines.
\end{itemize}

These findings not only contribute to our understanding of bias in human decision-making but also underscore the potential of advanced algorithms to foster collective intelligence from human inputs and mitigate bias. The following sections provide a detailed examination of these results, highlighting the role machine intelligence can play in enhancing decision-making processes.

\section{Results}
\subsection{Influence of Aggregation Algorithms on Collective Decision-Making Processes}\label{sec:alg_performance}

Here we compare the performance of various aggregation algorithms: selecting a participant at random (random expert), a Weighted Majority Vote (WMV), EXP4 \cite{auer2002nonstochastic}, MetaCMAB \cite{abels2023dealing}, and ExpertiseTree \cite{pmlr-v202-abels23a}. Each approach is included in the study for its unique contribution to understanding and improving collective decision-making. We provide a description of these algorithms, as well as a more detailed justification for their inclusion in the methods section. We focus first on how these algorithms influence the overall decision quality compared to individual participants.

\paragraph{Accuracy}
\begin{figure*} 
\includegraphics[width=\textwidth]{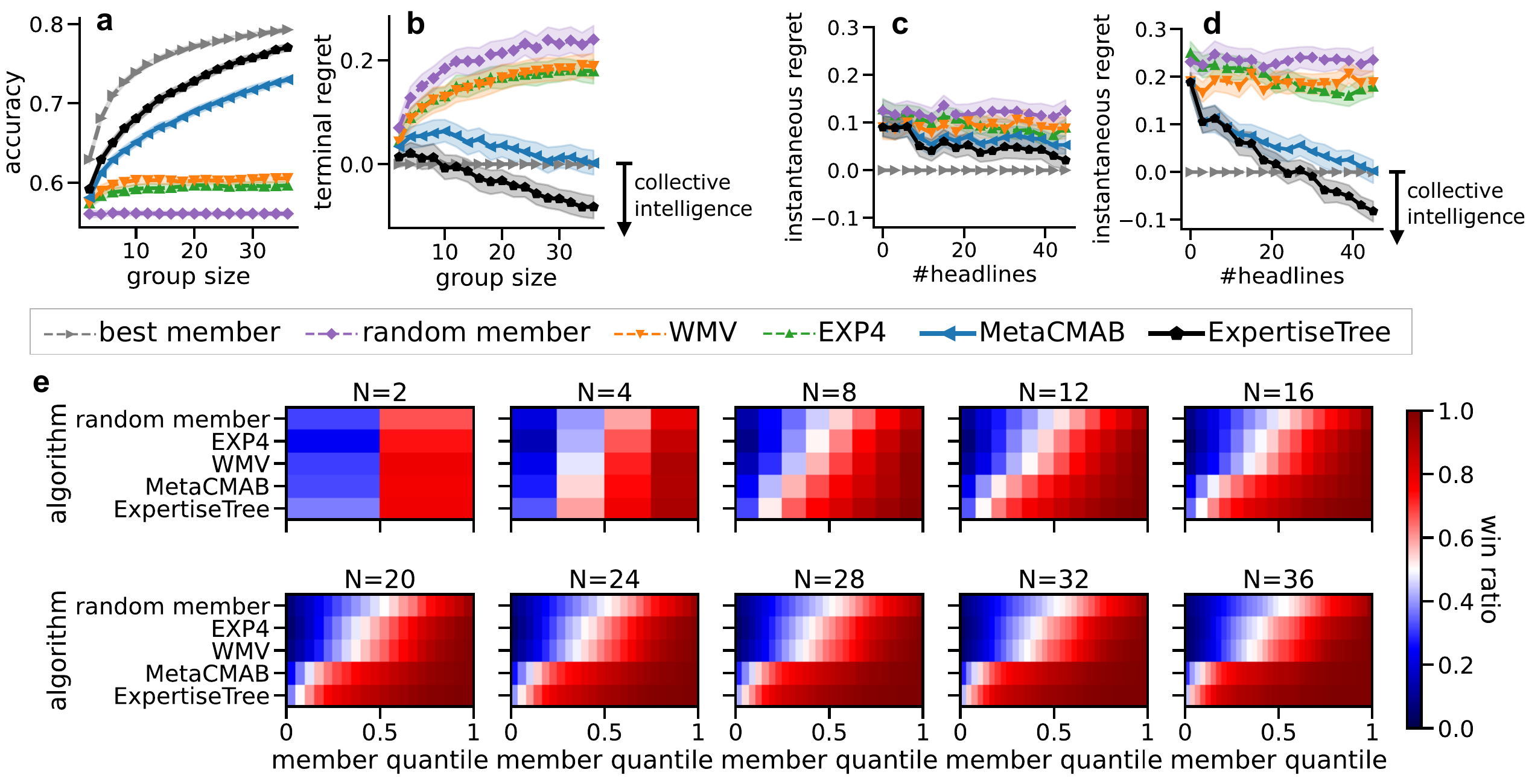}
    \caption{Performance as a function of the number of participants. Shaded areas around the lines in panels \textbf{a-d} represent $95\%$ confidence intervals. \textbf{a}, displays the accuracy (i.e., the proportion of correctly identified headlines, see Methods) of each algorithm for each group size. \textbf{b}, shows, for each group size, the terminal instantaneous regret (see Methods): the difference in performance between the best group member and the algorithm for the final decision. Negative values in \textbf{b} show when the collective outperforms the single best group member.
    Panels \textbf{c} and \textbf{d}, depict for, respectively, groups of size $4$ and $36$ the instantaneous regret as learning progresses --- the difference in performance between the best group member and the algorithm for the decision at that time (see Methods). These two plots show at which point algorithms surpass the single best group member. \textbf{e}, presents the algorithmic improvement on expertise for different algorithms and different group sizes ($N$). Within each heatmap, group members are ranked by accuracy, and the x-axis indicates the member's performance quantiles. In particular, the lower the quantile, the more accurate the group member. For each row within a heatmap, each cell represents the proportion of cases for which the respective algorithm surpasses the corresponding group member's performance. In particular, the left-most cell in each diagram corresponds to the best group member. Non-zero values for this cell suggest the emergence of collective intelligence.
    }
    \label{fig:performance_plots}
\end{figure*}
\letterref{fig:performance_plots}{a} shows overall accuracy as a function of group size for these algorithms and compares it to the best group member (identified in hindsight). \letterref{fig:performance_plots}{b} gives terminal regret values, i.e., the difference in performance with the best expert for the last round. Based on Wilcoxon tests (see Methods) with an $\alpha=0.05$ significance threshold, the performance of all methods is significantly different, except for the terminal regret of EXP4 and the WMV. 

\paragraph{Round-by-Round Analysis of Instantaneous Regret}

To further analyze these dynamics, we analyzed the round-by-round performance for group sizes of 4 and 36 participants, specifically focusing on instantaneous regret (\letterref{fig:performance_plots}{c-d}). This metric measures the performance gap between each algorithm and the best performing group member at specific time steps. In particular, a negative regret indicates performance surpassing the single best group member, indicative of the emergence of collective intelligence. As it collects more experiences, ExpertiseTree approaches this threshold, and ultimately surpasses it for larger groups (compare $N=4$ and $N=36$, \letterref{fig:performance_plots}{c-d}).

\paragraph{Win Percentage}
An alternative way of evaluating the prevalence of collective intelligence is by measuring how often collective decision-making algorithms surpass the performance of the single best group member. \letterref{fig:performance_plots}{e} provides this comprehensive view of how different algorithms fare against individual participants. While all methods tend to outperform a majority of group members, MetaCMAB and the ExpertiseTree approach consistently outperform all but 2 (for MetaCMAB) or 1 (for ExpertiseTree) group members. For example, for group size $N=36$, ExpertiseTree exceeds the performance of the top group member in $45\%$ of cases (as shown in the bottom left cell of the relevant heatmap), and surpasses the second-best member in $56\%$ of simulations (second leftmost cell). On the other hand, MetaCMAB generally matches the performance of the second-best member (with a win ratio of $0.40$) and less often outperforms the top member, achieving a win ratio of $0.3$.

\subsection{Dissecting Participant Biases}
\label{sec:performance_by_sentiment}

This section explores the biases exhibited by participants within our collective decision-making framework. Through this analysis, we aim to provide a foundation for understanding to what extent collective decision-making systems, as discussed in \autoref{sec:cdm_bias_mitigation}, can mitigate biases.

\paragraph{Demographic Differences in Performance}\label{sec:demographic}

We first investigate whether the participants' demographics correlate with differences in performance across headline categories and find the following, as illustrated by \autoref{fig:perf_by_dem}:

\begin{figure*}
\centering
\includegraphics[height=0.15\textwidth]{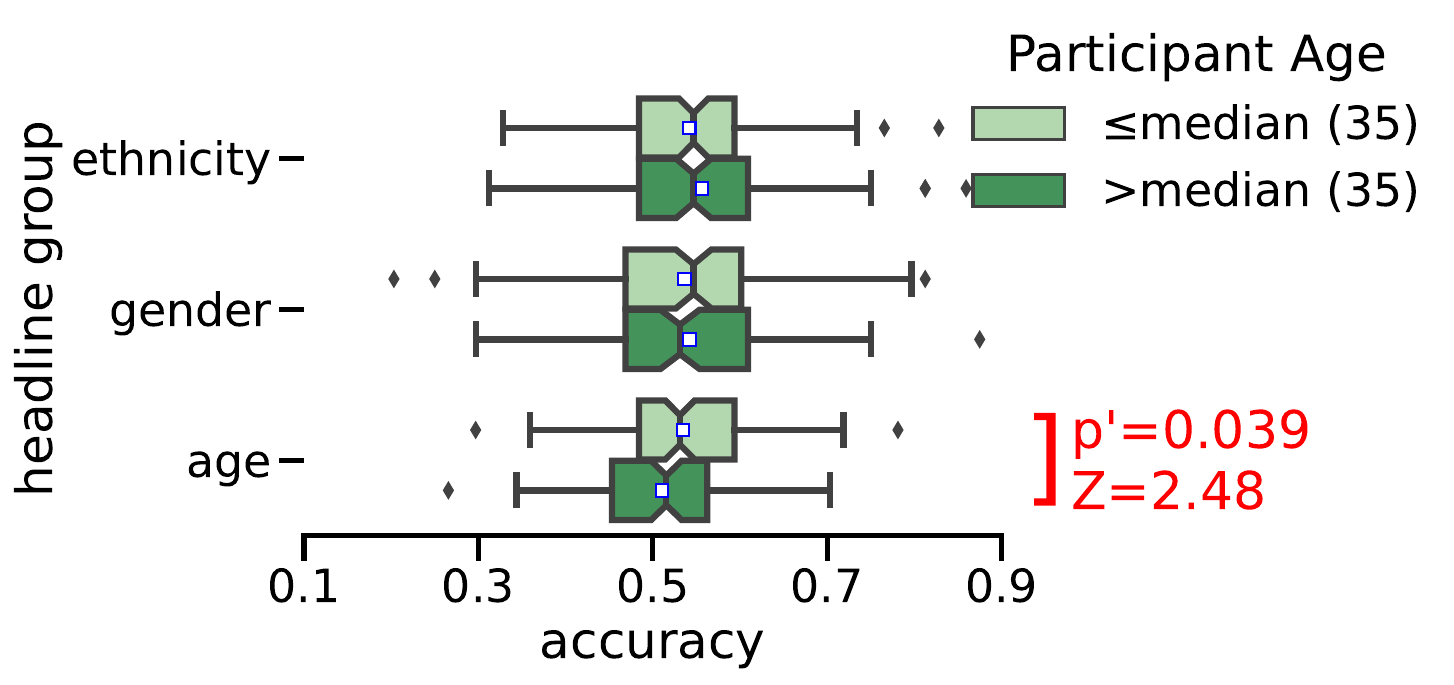}
\includegraphics[height=0.15\textwidth]{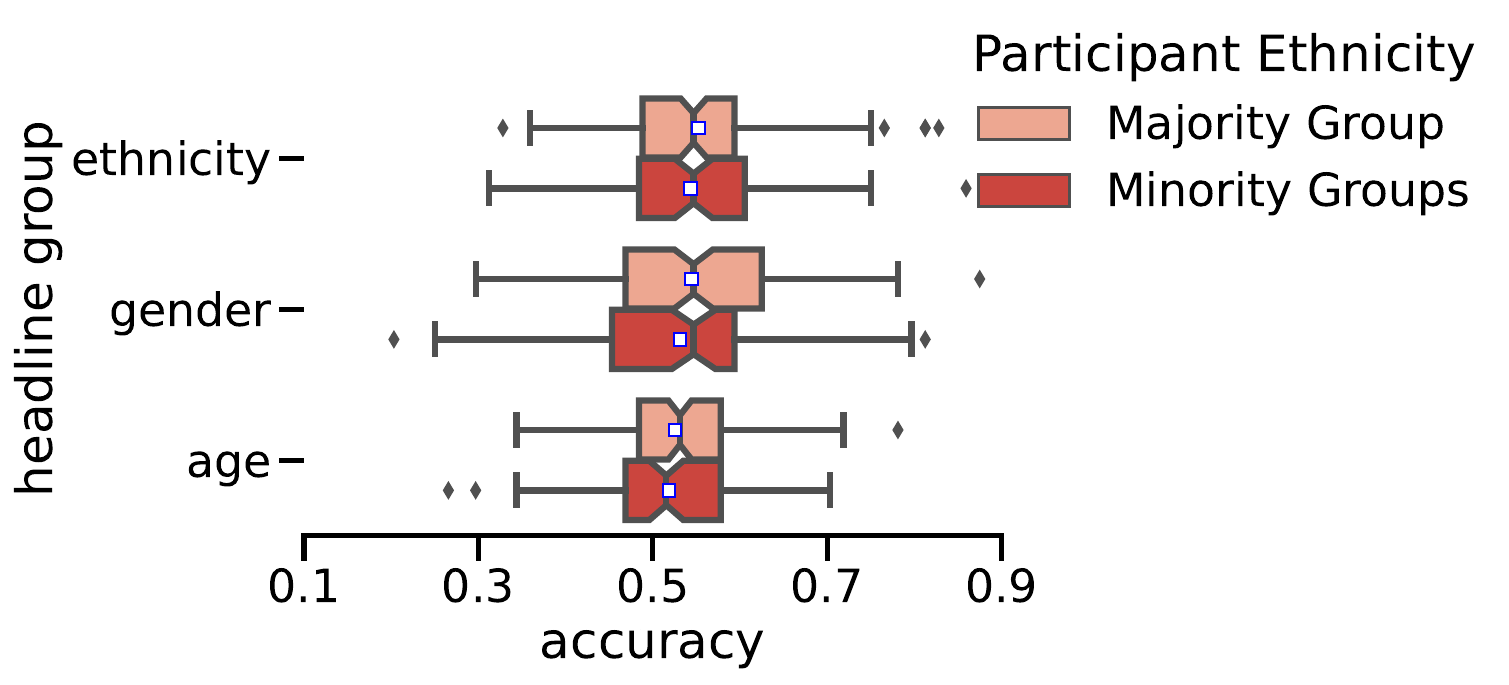}
\includegraphics[height=0.15\textwidth]  {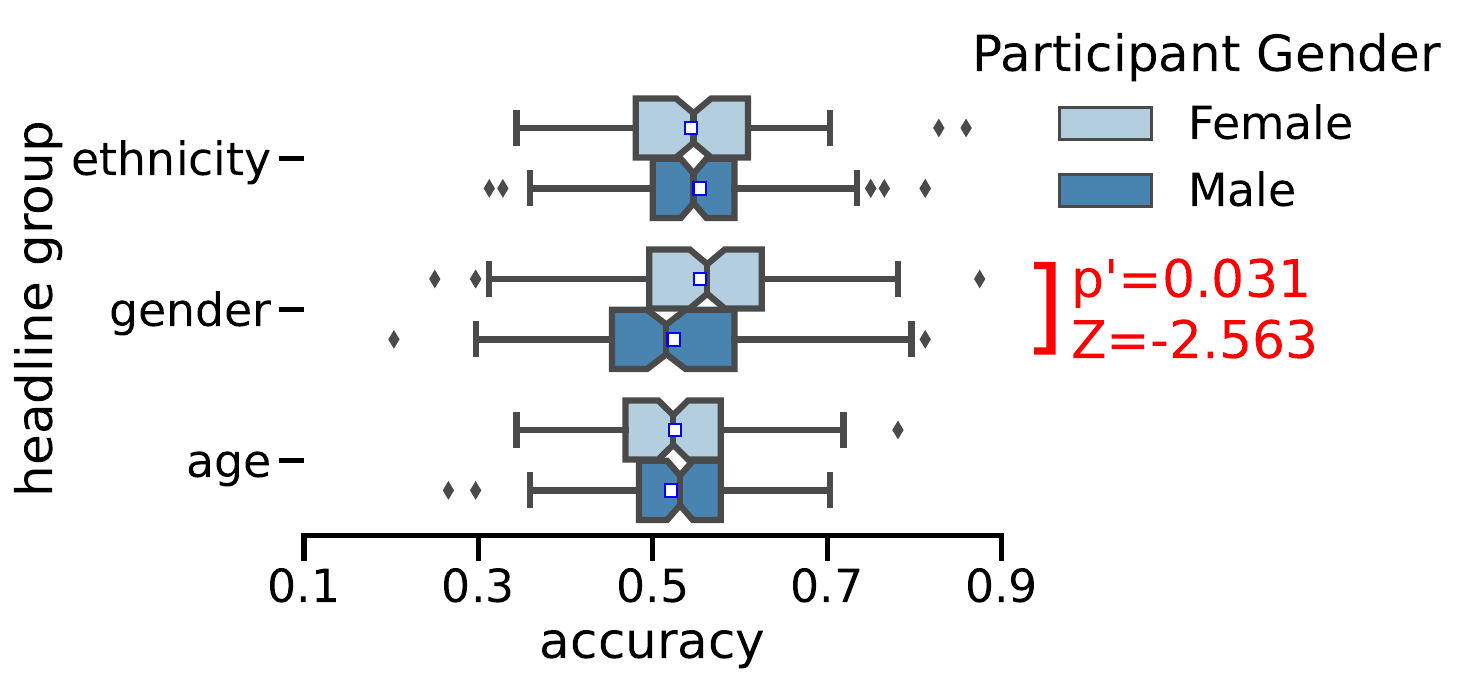}
\caption{Participants' accuracy as a function of their demographic information and the type of headline they respond to. The figures, from left to right, depict the accuracy in responding to each headline category in function of three demographic variables: age group (threshold at the median participant age of $35$), ethnicity (partitioned into majority and minority ethnicities, see \autoref{fig:demographic_histplots}), and gender. The GEEs fitting reveals statistically significant differences (Bonferroni adjusted p-values $p'<0.05$) in age and gender related questions. Specifically, older participants showed slightly lower accuracy in responding to age-related questions, and male participants exhibited lower accuracy in responding to questions related to gender. }
\label{fig:perf_by_dem}
\end{figure*}

\begin{itemize} 
    \item There is a significant difference in accuracy between men and women, especially when responding to gender-related headlines. Men exhibited a lower average accuracy ($0.524$) compared to women ($0.554$), with a statistically significant difference ($\beta = 0.0303, p= 0.01, ci=[-0.053,-0.007]$).
    \item Participants' performance varied significantly with age, particularly when interpreting headlines about different age groups. Individuals younger than 35 years were more accurate in identifying altered headlines about age ($0.535$ accuracy) compared to those older than 35 years ($0.511$ accuracy), with a significant difference ($\beta = 0.024, p = 0.013, ci=[0.005,0.043]$). 
    \item Unlike gender and age, we did not observe significant differences in accuracy based on participants' ethnic backgrounds. Both majority and minority ethnic groups performed similarly across various headline types ($|\beta| \leq 0.013, p\geq 0.384$). 
\end{itemize}

\paragraph{Framing Effect}\label{par:framing_effect}
We next investigated the framing effect --- how the presentation of information shapes responses \citep{plous1993psychology} --- by comparing participant reactions to original versus altered headlines.
We hypothesize that, in the absence of framing effects, if a headline is believed to be true, its altered version would likely be deemed false, and vice versa. Deviations from this expectation indicate a framing effect, suggesting that people respond to more than just the headline's content.

\begin{figure*}
\centering
\includegraphics[width=1\textwidth]{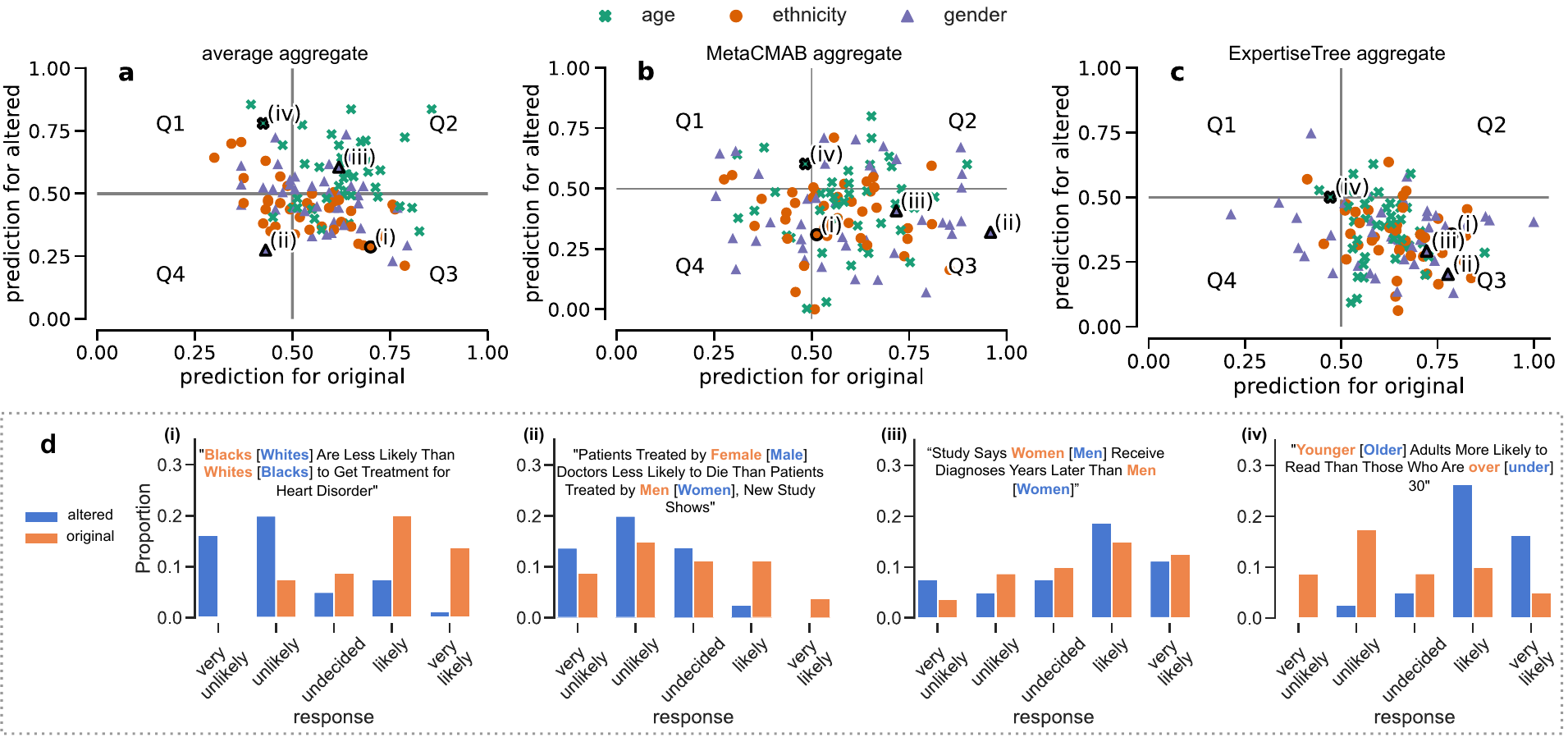}
\caption{Effects of framing in participants' responses. \textbf{a-c}, the quadrants represent false stereotypes (Q1), positive framing effects (Q2), common knowledge (Q3), and negative framing effects (Q4).
\textbf{a}, Each point represents a distinct headline; the x-coordinate displays the average response to its original form, and the y-coordinate shows the average response to its altered form.
\textbf{b}, The x and y coordinates indicate MetaCMAB's model predictions for the original and altered headlines, respectively.
\textbf{c}, In a similar manner, x and y coordinates reveal the predictions of ExpertiseTree's model for the original and altered headlines, respectively.
\textbf{d}, Response distributions for highlighted points (i), (ii), (iii) and (iv) in panels \textbf{a-c} are given as histograms. Each point's headline and response distribution are given in its original (orange), and [altered] (blue) form.}
\label{fig:scatter_real_altered}
\end{figure*}

Using the Mann-Whitney U test, we tested whether the response distribution for a real headline differed from its altered counterpart. A total of $44\%$ of headlines were found to induce a framing effect ($p<0.05$). To further dissect this phenomenon, \autoref{fig:scatter_real_altered} illustrates the relationship between average responses to original and altered headlines.

\paragraph{Group Biases}\label{par:disparity}

Next, we contrasted responses between groups within the same category. For example, are responses to headlines reporting positive outcomes for men different from those reporting positive outcomes for women?\footnote{In our experimental design, any negative outcome for one group was implicitly a positive outcome for the complementary group. Thus, a positive outcome for men is implicitly a negative outcome for women, and conversely, a positive outcome for women is implicitly a negative outcome for men} 

\begin{figure}
    \centering
    
    \begin{subfigure}[b]{.4\textwidth}
        \centering
        \includegraphics[width=1\textwidth]{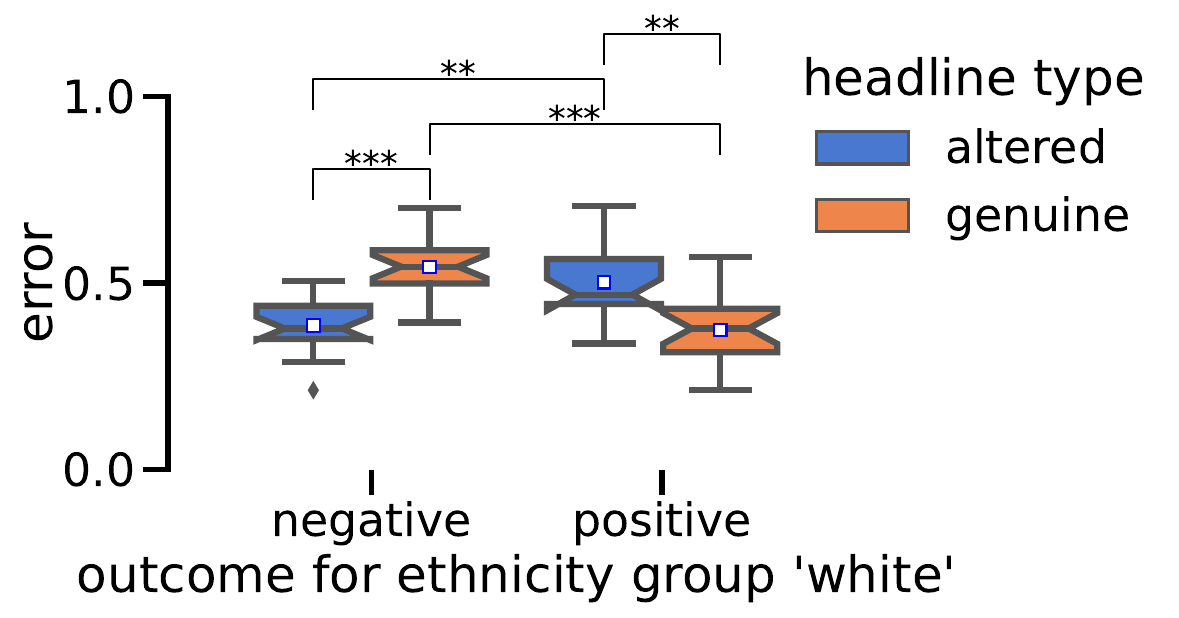}
        \caption{Headlines involving ethnicity}
        \label{fig:error_by_ethnicity_sentiment}
    \end{subfigure}
    \begin{subfigure}[b]{.4\textwidth}
        \centering
        \includegraphics[width=1\textwidth]{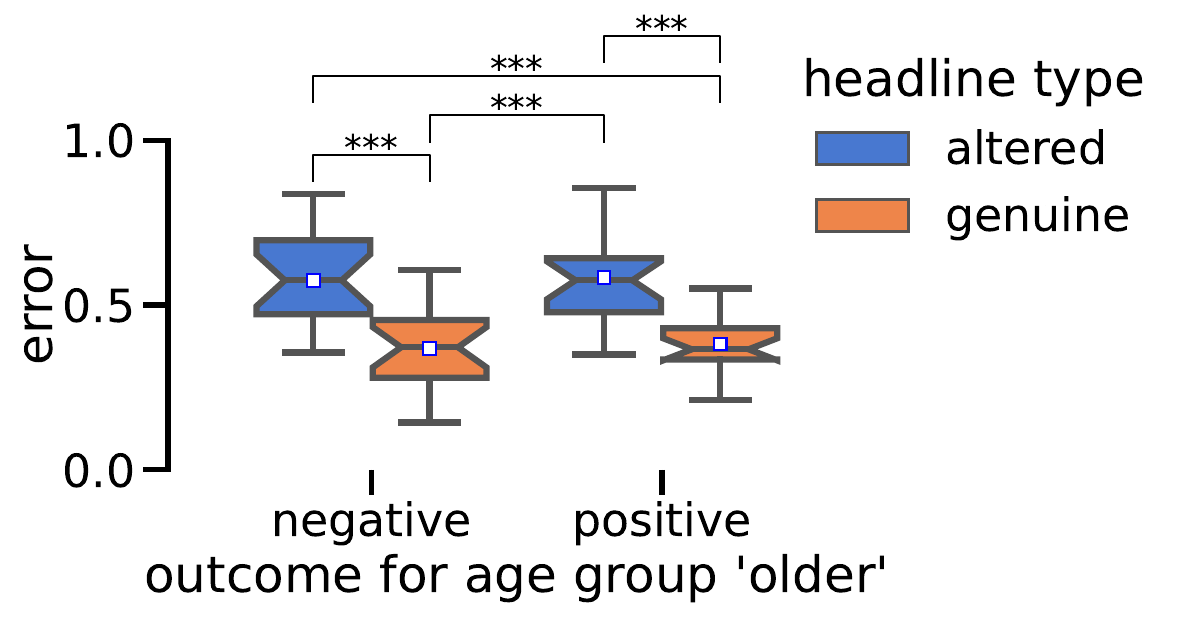}
        \caption{Headlines involving age}
        \label{fig:error_by_age_sentiment}
    \end{subfigure}
    \begin{subfigure}[b]{.4\textwidth}
        \centering
        \includegraphics[width=1\textwidth]{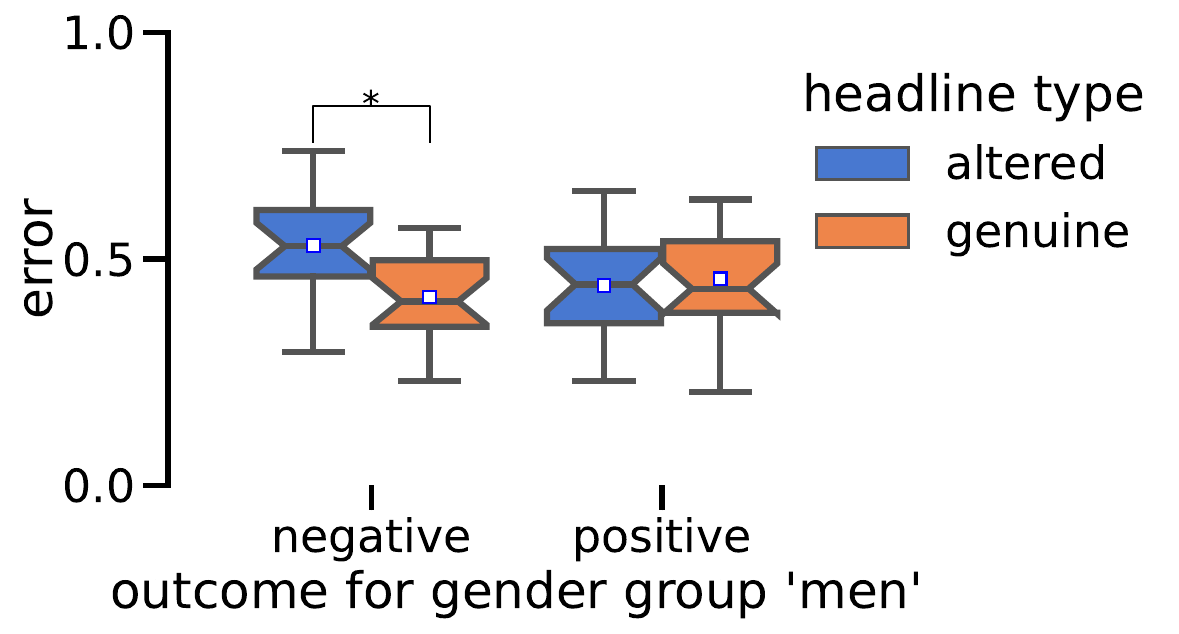}
        \caption{Headlines involving gender}
        \label{fig:error_by_gender_sentiment}
    \end{subfigure}
    \caption{Error rates categorized by sensitive attributes in headlines. Brackets indicate that there are significant differences (*, **, and *** indicating respectively p-values below $0.05$,$0.01$, and $0.001$) between pairs of boxplots as identified by Kruskal-Wallis H-tests (ethnicity: $H=33.995$, $p<0.001$, $df=3$, $\eta^2=0.408$, age: $H=34.167$, $p<0.001$, $df=3$, $\eta^2=0.410$ gender: $H=10.421$, $p=0.0153$, $df=3$, $\eta^2=0.098$) followed by Dunn's tests (see Methods).
    }
    \label{fig:error_by_sentiment}
\end{figure}

\autoref{fig:error_by_ethnicity_sentiment} shows that participants are more prone to errors when evaluating altered headlines which report positive outcomes for white people, and real headlines which report negative outcomes for white people. 

A similar analysis as a function of age, shown in \autoref{fig:error_by_age_sentiment} reveals that the disparity in error between real and altered headlines is consistent across sentiments. In particular, it shows that participants are strong at identifying real headlines, but weak at identifying altered headlines.

Finally, the analysis by gender, given in \autoref{fig:error_by_gender_sentiment} reveals that errors are relatively consistent across all but one group; headlines altered to be negative for Males.

\subsection{Collective Decision-Making Models Mitigate Biases}\label{sec:cdm_bias_mitigation}
In this section we compare and contrast the prevalence of biases in aggregations through either an average or adaptive methods MetaCMAB and ExpertiseTree. We restrict the analysis of EXP4 to the Supplementary \autoref{sec:framing_exp} as we found it did not improve over the performance of a simple average.

\paragraph{Reduced Framing Effects}
Our hypothesis posits that advanced collective decision-making algorithms, such as MetaCMAB and ExpertiseTree, should show a notable reduction in framing effects. This would manifest as decreased densities in Quadrants Q1 (False Stereotypes), Q2 (Positive Framing Effect), and Q4 (Negative Framing Effect) of the framing effect analysis.

We evaluated this hypothesis by analyzing the aggregated predictions of these models (see \letterref{fig:scatter_real_altered}{b} for MetaCMAB and \letterref{fig:scatter_real_altered}{c} for ExpertiseTree). These plots indicate a reduced proportion of headlines in Q1 ($19\rightarrow 8$) and an increased proportion of headlines in Q4 ($15 \rightarrow 25$) for MetaCMAB, as well as a reduced proportion of headlines in Q1 ($19 \rightarrow 4$) and Q2 ($34 \rightarrow 11$) and an increased proportion in Q3 ($52\rightarrow 97$) for ExpertiseTree.

\paragraph{Reduced Group Biases}\label{sec:group_biases}
We now evaluate whether different algorithms are predictive of different error rates conditioned on the category. This evaluation involves comparing error rates from three sources: average participant predictions, MetaCMAB predictions, and ExpertiseTree predictions. We summarize here the results and provide full GEE tables in the Supplementary Information.

The GEE analysis for averaged responses highlights significant interactions across all variables with respect to prediction error. The baseline error (intercept) was estimated at $\beta = 0.579$ ($p < 0.001$, $ci=[0.536, 0.623]$). Negative effects were observed for headlines classified by ethnicity ($\beta = -0.135$, $p < 0.001$, $ci=[-0.204, -0.067]$) and gender ($\beta = -0.093$, $p = 0.049$,  $ci=[-0.186, -0.001]$), with significant interactions noted between headline class and alteration status for both categories.

For MetaCMAB's predictions, the extent of bias was reduced. The baseline error was lower ($\beta = 0.431$, $p < 0.001$, $ci=[0.377509, 0.484283]$), and while the ethnicity headline class maintained a significant effect ($\beta = -0.087$, $p = 0.021$, $ci=[-0.160649, -0.012963]$), the gender effect was not significant. MetaCMAB's error is however significantly worse for genuine headlines than for altered headlines ($\beta = -0.079$, $p=0.023$, $ci=[-0.148, -0.011]$). The interaction effect of genuine headlines also remained significant in relation to ethnicity ($\beta = 0.132$, $p < 0.001$, $ci=[0.088,  0.179]$).

In contrast, the GEE analysis for ExpertiseTree's predictions showed that none of the variables had a significant effect on the prediction errors. The intercept was $\beta = 0.319$ ($p < 0.001$, $ci=[0.287, 0.351]$), and all other parameters concerning headline classes (both ethnicity and gender) and their interaction with altered status were non-significant (e.g., ethnicity headline class: $\beta = -0.053$, $p = 0.108$, $ci=[-0.118 , 0.0117]$).

\begin{figure}
    \centering
    \includegraphics[width=.35\textwidth]{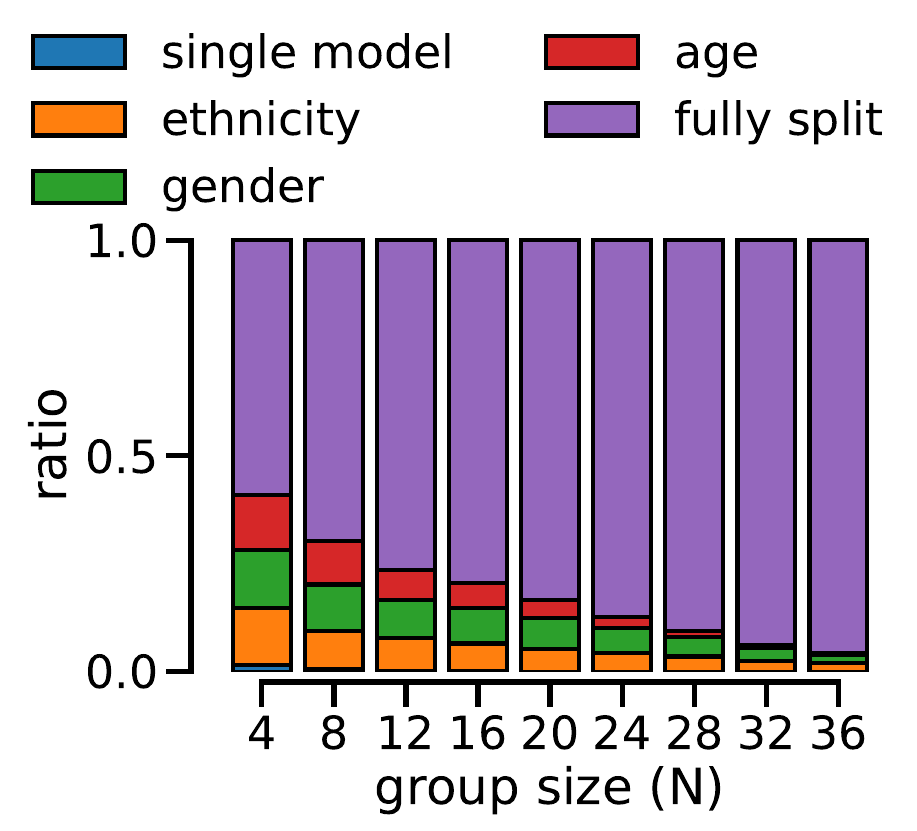}
    \caption{Prevalence of splits learned by ExpertiseTree as a function of the number of members in a group. ExpertiseTree can learn to either i) not split, ii) split off the ethnicity category, iii) split off the gender category, iv) split off the age category, v) split all categories. For (ii) --- and analogously for (iii) and (iv) ---  this implies that one model is learned for all headlines relating to ethnicity, while another model is learned for both the gender and age category.}
    \label{fig:model_distributions}
\end{figure}

 A comparison of the splits learned by ExpertiseTree (see \autoref{fig:model_distributions}) shows its trees grow deeper as the group size increases. In particular, for large group sizes the ExpertiseTree learns distinct models for all headline categories.

\section{Discussion}
Our investigation into collective decision-making, biases, and the effectiveness of adaptive methods, reveals several key insights with implications for the field of machine intelligence.

\paragraph{Collective Intelligence through Online Machine Learning} With no prior knowledge of the expertise of the group members, efficient collective decision-making systems need to learn to surpass the performance of the single best group member. When group members provide honest confidence estimates, these can be used to weigh their input. In our experiment, participants were given the option to express strong confidence (e.g., 'very likely'), weak confidence (e.g., 'likely'), or a lack of knowledge ('undecided'). If participant responses were properly calibrated, i.e., their expressed confidence aligned with their expected accuracy, previous work indicates that the resulting confidence-weighted majority vote (WMV) would consistently outperform the single best group member \cite{Grofman1983}. As \letterref{fig:performance_plots}{a} illustrates, however, this is not the case: the WMV consistently performs worse than the single best group member. 

To tackle this confidence bias, various online machine learning approaches have been proposed to supplement participants' self-reported confidence with algorithmically determined weights. These weights are iteratively refined as more information regarding the participants' accuracy is gathered, thereby enhancing the system's ability to adjust its aggregation over time.

We compared here three different approaches: EXP4, MetaCMAB, and ExpertiseTree.
We find that MetaCMAB and ExpertiseTree, which act on an aggregate over all group members, perform significantly better than EXP4, which is designed to select the single best decision-maker in a group. In fact, as the group sizes increase, their accuracy converges towards the level achieved by the best-performing group member. 

While ExpertiseTree typically does not outperform the top individual participant overall, its potential becomes more evident in later stages of the decision-making process. As shown in \letterref{fig:performance_plots}{b}, this algorithm ultimately surpass the best participant's performance in larger groups ($N\geq 8$). Therefore, although initial uncertainty limits overall performance, these results suggest that this algorithm eventually facilitate the emergence of collective intelligence. In contrast, while a WMV can achieve collective intelligence in specific scenarios (homogeneous performance levels and low correlation between group members, or properly calibrated confidence, see \cite{abels2023dealing}), participant responses in this context do not satisfy these necessary conditions. 

Results in terms of win percentages (\letterref{fig:performance_plots}{e}) confirm that the WMV and EXP4 algorithms consistently underperform in comparison with the top-performing group member. In contrast, algorithms emphasizing collective intelligence (MetaCMAB and ExpertiseTree) frequently outperform the single best group member.

Overall, our findings align with prior results on the performance of these algorithms established on synthetic expertise: the use of adaptive algorithms enables a dynamic and responsive collective decision-making process. Although they initially are on par with a WMV, MetaCMAB and ExpertiseTree quickly improve their performance by adjusting the importance assigned to group members. Notably, for groups of 36, MetaCMAB and ExpertiseTree begin to surpass the best expert after 42 and 24 headlines, respectively. This demonstrates that these algorithms quickly adapt to the collective pool of knowledge, often surpassing the performance of the most skilled individual in larger groups. This collective intelligence is achieved by leveraging diverse inputs without prior knowledge about participants' expertise.

\paragraph{Headline Categories and Implicit Biases} 

Based on prior research \cite{greenwald2006implicit,craig2016stigma}, we expected to observe differences in participants' responses to various headline categories, such as gender, ethnicity, or age. For instance, individuals might be less inclined to question stereotypes concerning age groups but more cautious when it comes to stereotypes associated with ethnic groups. 
Our analysis of how participants respond to different headline categories yielded notable trends in skepticism. In particular, the quadrants of \letterref{fig:scatter_real_altered}{a} align with four types of beliefs: 
\begin{itemize}
    \item (Q1, False Stereotypes)  Stereotypes skew judgment, leading to consistent misjudgment of these headlines that align with preconceived notions.
    \item (Q2, Positive Framing Effect)  Headlines are perceived as true regardless of alterations, indicating a lower skepticism induced by their phrasing or content.
    \item (Q3, Common Knowledge) Headlines matching widespread beliefs are accurately judged, irrespective of their alteration.
    \item (Q4, Negative Framing Effect) Converse of Q2; these headlines are consistently deemed false, pointing towards inherent skepticism or bias in their presentation.
\end{itemize}

Interestingly, headlines related to age were predominantly found in Q2 and rarely in Q4. On the contrary, ethnicity-related headlines were absent from Q2. This suggests participants generally exhibited reduced skepticism towards age-related headlines, contrasting with a heightened skepticism towards ethnic headlines. This pattern persisted even when headlines were altered to focus on different sensitive groups (see Supplementary Information); responses for a headline whose subject is changed from an ethnic group to an age group will be less skeptical. This suggests that this skepticism (or lack thereof) is not a feature of the specific headlines, but rather of the sensitive group it contains. This finding aligns with research by \cite{nosek2007pervasiveness}, suggesting that people are more inclined to associate age with positive or negative concepts than ethnicity. This could explain the greater likelihood of believing age-related headlines. 

When detailed by subcategory, our results (in particular \autoref{fig:error_by_ethnicity_sentiment}) suggest  that participants in the experiments were more skeptical of headlines which report negative outcomes for white people.  

These patterns can also be understood in terms of confirmation bias, the tendency for people to favor information that confirms their pre-existing beliefs or biases \cite{oswald2004confirmation}. This phenomenon, notably discussed in studies of crowdsourcing  \citep{gemalmaz2021accounting,draws2021checklist}, offers further insights into participants' responses. For example, if individuals hold implicit biases that view certain groups in a specific light, they are more likely to believe headlines that confirm these biases, and dismiss those that challenge them. 

\paragraph{Demographic Differences in Performance} 

Results on the Implicit Association Test (IAT) \citep{greenwald2006implicit} suggest that most groups, with the exception of African-Americans, displayed an implicit bias against African-Americans. Similarly, \cite{craig2016stigma} found stigmatized groups can form coalitions, in the sense that members of a stigmatized group might be more sensitive to the discrimination of other stigmatized groups. We found that performance disparities between demographic groups predominantly appeared in dimensions relevant to those groups (see \autoref{fig:perf_by_dem}). For instance, differences in gender or age influenced the accuracy of responses to headlines pertaining to those specific categories. This suggests that personal identity factors play a crucial role in shaping perceptions and biases.

\paragraph{Mitigating Bias in Collective Decision-Making through Machine Intelligence}

Considering the enhancements in performance shown by MetaCMAB and ExpertiseTree compared to individual expert predictions, we investigated if these improvements also indicate a reduction in group bias, i.e., whether the group (e.g., gender, ethnicity, or age) for which a prediction is given impacts the predictor's error. 

Our analysis reveals that individuals often display biased performance. When individual responses are pooled through an average, these biases tend to permeate, resulting in biased aggregations. In contrast, adaptive aggregations acquired through for example MetaCMAB or ExpertiseTree should mitigate these biases. 

We found that, unlike other methods, ExpertiseTree's performance remained was not significantly affected by changes in headline categories, suggesting individual biases were effectively mitigated.

What is more, the distribution of points in \letterref{fig:scatter_real_altered}{b} suggests a decrease in false stereotypes and a reduced positive bias for MetaCMAB, but a similar prevalence of negative framing effects. In contrast, we found less populated biased quadrants for ExpertiseTree. This supports its robustness in mitigating false stereotypes and framing effects, making it a valuable tool in creating unbiased, balanced decision-making processes.

 Finally, the prevalence of fully grown trees for larger group sizes (see \autoref{fig:model_distributions}) suggests that when there are more group members available, the ExpertiseTree algorithm is more likely to find a subset of members which is stronger for a given headline category.

\section{Conclusion}
In this study, we explored the impact of individual and group biases in collective decision-making, particularly in the context of identifying the authenticity of news headlines. Our findings underscore the potential of adaptive aggregation algorithms, such as MetaCMAB and ExpertiseTree, in enhancing collective intelligence and mitigating biases.

We demonstrated that while individual responses are prone to various biases, including those related to demographic factors, adaptive algorithms can effectively counteract these biases. The ExpertiseTree algorithm, in particular, showed a significant reduction in framing effects and group biases compared to individual responses or simple aggregation methods like the weighted majority vote.

Those findings contribute to the understanding of how machine learning algorithms can be utilized to improve the quality of collective decisions. By integrating a diverse set of individual opinions and dynamically adjusting to the varying levels of expertise, these algorithms enable the emergence of collective intelligence, which is crucial in contexts where decision-making involves complex, sensitive issues.

In summary, our research underscores the importance of understanding and addressing biases in human advice to enhance the performance of collective decision-making. The use of advanced machine learning algorithms, such as MetaCMAB and ExpertiseTree, offers a promising avenue for achieving this goal. By effectively aggregating diverse human opinions and dynamically adjusting to various levels of expertise and bias, these algorithms pave the way for collectively intelligent, fair, and effective decision-making processes.

Finally, it is worth noting that complementing the online learning method we propose with proactive bias reduction strategies could further enhance overall bias mitigation. In contrast to our approach, various studies have proposed to mitigate biases by promoting deeper analytical thinking \citep{strack2004reflective,martire2014interpretation,croskerry2013cognitive,lau2009can,zizzo2000violation}. Techniques such as prompting participants to justify their decisions or encouraging them to reconsider their initial responses have been effective in reducing biases  \citep{vieider2009effect,spengler1995scientist,bago2020fake,parmley2006effects}. Although our primary goal was to optimize the extraction of knowledge from collected responses, future work could study how well these additional approaches synergize with ours.

\section{Methods}
We developed an online experiment aimed at gathering human decisions regarding fake news\footnote{This experiment has been preregistered \citep{osfreg}, we compare this paper to the preregistration in ``Preregistration Discussion" section below.}. 

To focus on individual biases, we designed our experiment without any interaction between participants, ensuring that the collected responses were independent. Following this, we sampled subgroups from the participant pool to evaluate the performance of the different online learning algorithms.

\subsection{Questionnaire}

In order to ground our results in a realistic problem, we presented participants with a fake news problem. In doing so, we are emulating the use of human fact checkers as a tool to combat the spread of fake news on social media. Our aim is thus to evaluate how laypeople perform at this task, what factors influence their choices, and how they could be enhanced by a collective decision-making system. Participants were presented with a series of headlines for which they were asked to express a likelihood of the headline being real. In order to elicit possibly biased responses, we focused on headlines which involved sensitive groups. 

In particular, we collected headlines from \citep{DVN/SYBGZL_2018,mazumder2014news,Medscape} based on the following criteria:
\begin{itemize}
 \item  The headline (implicitly or explicitly) contrasts two sensitive groups (e.g., \textit{``Men more likely than women to say they are financially better off since last year"})
 \item The headline should present a clear negative or positive outcome. For example, it is not clear whether \textit{``Poll: Kanye more popular with whites than nonwhites"} is positive or negative, but \textit{``African-Americans, Hispanics, dying at faster rate of fentanyl overdoses than whites: analysis"} is clearly a negative outcome for African-Americans and Hispanics.
\end{itemize}
Participants were presented with a mix of such headlines: $50\%$ were unaltered, while the other $50\%$ had their sensitive group swapped. 

Alterations were made by swapping sensitive groups with their complementary group. For example, ``Men" $\leftrightarrow$ ``Women", ``Older people" $\leftrightarrow$ ``Younger people", or ``African-Americans" $\leftrightarrow$ ``White Americans".  

Each headline was characterized by a sensitive group (thus whether it concerned a gender group, an ethnic group, or an age group), by a sentiment (positive or negative), and by a truth value (whether the headline was real or altered). 

Headlines for each participant were selected to represent a balanced mix of these features. The number of headlines required to achieve this balance can be computed as:
\begin{multline*}
(\text{\# of sensitive groups} \times \text{\# of sentiments})\\ 
\times (\text{\# of outcomes} \times \text{\# of truth values})\\ 
= (3 \times 2) \times (2 \times 2) = 24.    
\end{multline*}
Each treatment, or set of questions, should therefore contain a multiple of $24$ headlines.

Moreover, to understand the influence of presentation, every headline was included in both its real and altered versions, distributed across different treatments.

Consequently, our finalized dataset encompasses $5$ sets of questions (=treatments), with each set containing $48$ headlines. These $240$ headlines contain $40$ instances for each of the 6 sensitive groups. By distributing these over $5$ treatments, we ensured each participant saw $8$ real and $8$ altered headlines for every sensitive group. Note that a positive sentiment towards one group implies a negative sentiment towards its complementary group. For example, \textit{``Across Age Groups, Whites Fared Worse in Employment Rates"} is both negative for whites and implicitly positive for African-Americans. We thus have $16$ real ($8$ positive and $8$ negative) and $16$ altered (again $8$ positive and $8$ negative) headlines per group. 

We sequentially presented the headlines to participants and asked them to assess the likelihood of each being real. For each headline, participants were given $5$ choices; \{``very unlikely", ``unlikely", ``undecided", ``likely", ``very likely"\}. 

\subsection{Participants and Data Collection}

\begin{figure*}
    \centering
    \includegraphics[width=.9\textwidth]{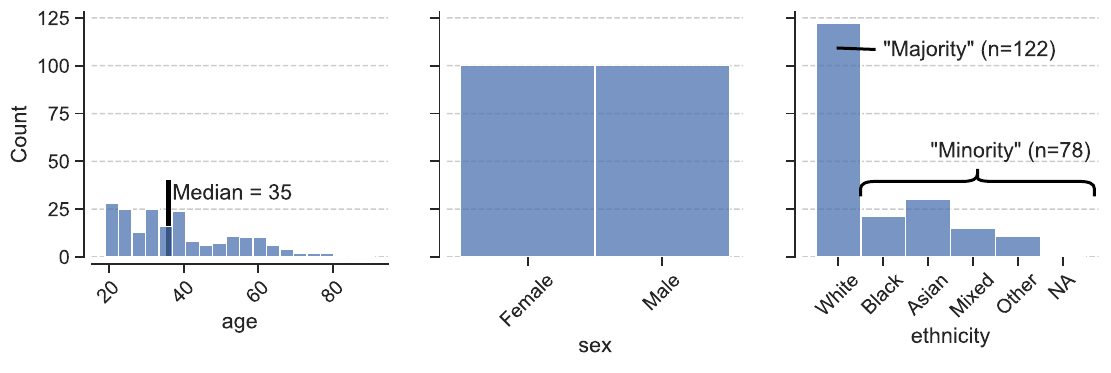}
    \caption{Demographic distribution of participants. Age categories are established by splitting participants according to the median age of $35$. Ethnicity categories, Majority and Minority, are assigned according to whether participants belong to the most frequent group (i.e., White). NA = Not Available}
    \label{fig:demographic_histplots}
\end{figure*}

We recruited 200\footnote{Additionally, 26 participants who did not complete the questionnaire were excluded from this analysis.} participants via the Prolific crowdsourcing platform \citep{palan2018prolific}. While the distribution of participants by gender was balanced, balancing participants by ethnicity or by age is more challenging through Prolific. We therefore did not impose restrictions on participation by ethnicity or by age, except that all participants are $\geq 18$ years old. As a result, participants were predominantly white ($N_{\text{white}}=122$, $N_{\text{non-white}}=78$, see \autoref{fig:demographic_histplots}), reflecting demographics of the sampled population.

Participants were paid a flat sum for participation at a rate of 6£ an hour. Median completion time was $8$ minutes $58$ seconds. Participants were uniformly assigned to one of the 5 treatments (=set of $48$ headlines). Only participants who completed the questionnaire were included in our analysis. Consequently, our dataset contains a total of $5 \times 40 \times 48 = 9600$ responses.

Along with responses, we captured the following metrics: the headline's sensitive group, the order in which responses were given, participants' demographic data (age, gender, and ethnicity), headline sentiment (positive or not), its authenticity (altered or not), and response time.

\subsection{Evaluating Performance}\label{sec:evaluating_performance}

To assess the quality of participants, we gauge the accuracy of their estimates against actual outcomes. Define \( p_n(h) \) as participant \( n \)'s response for headline \( h \). Following experimental evidence on how humans interpret these categories \citep{wintle2019verbal}, we map the responses to numerical quantities as follows: 
\[
p_n(h) = \begin{cases} 
0 & \text{if ``very unlikely"} \\
0.25 & \text{if ``unlikely"} \\
0.5 & \text{if ``undecided"} \\
0.75 & \text{if ``likely"} \\
1 & \text{if ``very likely"}
\end{cases}
\]
The true value of \( h \) is denoted as \( y(h) \), with \( y(h) = 0 \) for altered headlines and \( 1 \) otherwise. Participant error is then quantified in terms of absolute difference: \( \epsilon(h,n) = |p_n(h)-y(h)| \). Conversely, the accuracy is the complement of the error: $a(h,n) = 1 - \epsilon(h,n)$, and takes as value $1$ if participant $n$ predicts the headline's class with high confidence.

In this setting, bias can be assessed by observing whether participants are more likely to associate specific outcomes (either negative or positive) with certain groups. Additionally, differences in accuracy or error rates based on the sensitive group featured in the headline can also serve as indicators of bias. For example, participant errors for headlines reporting positive outcomes for white people could be significantly larger than those same participants' errors for headlines reporting negative outcomes for white people. Such differences are explored in \autoref{fig:error_by_sentiment} and its related discussion.

\subsection{Collective Decision-Making Process}\label{sec:CDM_process}
While it is typically hard to properly estimate the performance of participants a priori, online learning algorithms allow us to learn how to aggregate expertise as we encounter new cases. We outline this decision-making setting in \autoref{alg:CDM}.

\begin{algorithm*}
 \caption{Collective Decision-Making for Headline Selection}\label{alg:expert-learning}\label{alg:CDM}
 \begin{algorithmic}[1]
 \State Initialize learner (e.g., EXP4 \citep{auer2002nonstochastic}, MetaCMAB \citep{abels2023dealing}, or ExpertiseTree \citep{pmlr-v202-abels23a}) with initial aggregation policy 
 \For{$t = 1, 2, ..., T$}
 \State Participants observe the set of headlines for round $t$
 \State Each participant rates a headline on how likely it is to be real
 \State Aggregate their ratings to select a headline
 \State Collect reward of $r_t=0$ if chosen headline was altered, $1$ otherwise
 \State Use collected penalty to update aggregation policy for $t+1$
 \EndFor
 \end{algorithmic}
 \end{algorithm*}

In practice, we bootstrap, for every group size $N$ and for each treatment (set of headlines), $1000$ subsets of $N$ participants with which we simulate this collective decision-making process for each of the methods we now describe.

\paragraph*{Chosen Methods}
Several methods have been developed to improve decision-making with expert advice. Among them,  we select four pivotal approaches that span from fundamental to state-of-the-art, providing a comprehensive view of online learning algorithms for collective decision-making. These approaches include: a simple majority vote serving as a baseline for comparison; EXP4 \cite{auer2002nonstochastic}, which uses observed performance to gradually pinpoint the single best participant; MetaCMAB \cite{abels2023dealing}, an online learning approach that enables more collective decisions by integrating diverse opinions; and ExpertiseTree \cite{pmlr-v202-abels23a}, which tailors aggregation strategies to specific contexts.

As illustrated in \autoref{fig:problem-illustration}, a key differentiator among these methods is their strategy for assigning weight to participants' opinions, ranging from equal weighting to complex, performance- and context-based adjustments. Crucially, the more advanced approaches (MetaCMAB and ExpertiseTree) learn to weight participants based on their responses, not only relative to the truth, but also relative to each other. These methods not only prioritize contributions from the highest performers, resembling strategies that rely on the collective input of empirically proven subsets of participants \citep{mannes2014wisdom}, but they also incorporate mechanisms to offset the influence of highly polarized groups of experts. By recognizing and adjusting for common errors among participants, these approaches can reduce the negative effects of polarized groups, whose insights, though potentially insightful, overlap significantly with those of others \citep{abels2023dealing}. Accounting for correlation among participants in this way is somewhat similar to approaches which explicitly contrast individual's contributions with that of the group \citep{budescu2015identifying}. The main distinction between those alternative approaches and the ones we test in this study is the inclusion of exploration techniques which prevent premature reliance on a limited group of experts.

This selection of approaches is intentionally diverse, covering both traditional and state-of-the-art methods to highlight the spectrum of possibilities for aggregating advice. From the benchmark simplicity of majority voting to the nuanced, context-sensitive aggregation offered by ExpertiseTree, each method contributes uniquely to our understanding of effective decision-making in varied scenarios. A more detailed description of these chosen methods and the rationale behind their selection is provided in \autoref{sec:algorithms}.
 
\paragraph*{Measures of Collective Decision-Making Performance}
Because we consider an online learning setting, we evaluate the performance on two metrics. First, on overall accuracy, i.e., the proportion of rounds for which an optimal decision was made. And secondly, on final performance, i.e., the expected performance at the end of training. In addition, we can analyze how performance changes as a function of the number of participants. 

Given our objective of enhancing the probability of choosing authentic headlines, our goal becomes maximizing the \emph{average reward} defined as $\bar{\mathcal{R}}_T = \frac{1}{T}\sum_{t=1}^T r_t$, where $r_t=1$ if the chosen headline is genuine and $0$ otherwise. 

 We denote by $\bar{\mathcal{R}}_T^n$ the average reward we would obtain if we were to select participant $n$ and choose at each round the headline which they rate highest. Maximizing over the set of participants, we can in hindsight select the optimal participant as being $n^* = argmax_{n=1}^N \bar{\mathcal{R}}_T^n$. 

 To facilitate comparison to the single best participant, we can juxtapose an algorithm's performance against the best participant's performance. Specifically, for each round, we compute the instantaneous regret: 
\begin{equation}\label{eq:instantaneous_regret}
    R_t = r^{n^*}_t - r_t 
\end{equation} 

Of particular interest is the instantaneous regret at time $t=T$, i.e., $R_T$, which we refer to in the main text as terminal instantaneous regret.

If our performance is lower than the best participant, this metric averages out positively; otherwise, it is negative. Thus, observing negative regret suggests the presence of collective intelligence.

\subsection{Establishing Significance}

\subsubsection{Confidence Intervals}
In \autoref{fig:performance_plots}, the confidence intervals for the means are determined using a bootstrapping method \cite{davison1997bootstrap}. For each data point, we perform the following steps: We draw $1,000$ separate samples from the original set of $5,000$ simulation results. Each of these samples is of size $1,000$ and is drawn with replacement. For each of these $1,000$ samples, we calculate its average. These $1,000$ averages are then used to form the distribution of means. The 95\% confidence intervals are then established by identifying the values at the 2.5th and 97.5th percentiles of this distribution, corresponding to the lower and upper limits of the interval.

To more rigorously evaluate the significance of our results, we make use of the following two-sided tests.
\subsubsection{Wilcoxon Tests}
The Wilcoxon test is a non-parametric test used to compare two paired groups \citep{conover1999practical}. It is suitable for data that is not normally distributed and assesses whether the differences between paired observations are symmetric about zero. In our study, we applied the Wilcoxon test to evaluate the differences between algorithms ran on the same sets of participants. For example, we might compare the performance of EXP4 on participant groups $P_1,P_2,P_3$ to that of MetaCMAB on those same participant groups. The performance of both algorithms on any of those participant groups is therefore paired. 

\subsubsection{Generalized Estimating Equations}
Generalized Estimating Equations (GEE, \citep{liang1986longitudinal}) are an extension of the generalized linear model framework, designed to address the presence of correlated observations. GEEs are particularly useful for handling data characterized by repeated measures or clustered data, as it provides robust parameter estimates even when the correlation structure is misspecified. 

For instance, the significance of the results on demographic differences in performance (\autoref{sec:demographic}) is determined by GEE models with the dependent variable being the accuracy, the primary predictor being the participant's concerned sensitive attribute, e.g., their gender, and by clustering by participant groups.

\subsubsection{Mann-Whitney U Tests}
The Mann-Whitney U test is a non-parametric test used to compare the distributions of two independent samples \citep{mann1947test}. It is based on the ranks of the data rather than their actual values and is especially useful when the assumptions of the t-test (like normality) are not met. In our research, we utilized the Mann-Whitney U test to compare the distributions of two groups when no pairing was present. For example, we use it to compare the distribution of responses between headlines. This should provide a robust assessment of the differences between them without making stringent assumptions about the underlying data distribution.

\subsubsection{Kruskal-Wallis H-test}
The Kruskal-Wallis H-test is a non-parametric method used to compare three or more independent samples \citep{kruskal1952use}. It is an extension of the Mann-Whitney U test to multiple groups and tests the null hypothesis that all groups come from the same distribution. In our study, we employed the Kruskal-Wallis H-test when comparing the distributions of more than two independent groups. For instance, we might use it to assess the distribution of responses across multiple headline categories, aiming to determine if at least one headline category differs significantly from the others in terms of response distribution.

\subsubsection{Dunn's Test}
Following a significant Kruskal-Wallis H-test result, post-hoc pairwise comparisons are necessary to determine which groups differ from each other. Dunn's test is a non-parametric post-hoc method used to conduct these pairwise comparisons while adjusting for multiple comparisons \citep{dunn1964multiple}. It compares the differences in the sum of ranks between groups, similar to the Mann-Whitney U test but with a correction for multiple tests. In this paper, after finding a significant Kruskal-Wallis result, we applied Dunn's test to pinpoint which specific groups had significantly different distributions. For example, if we found a difference in response distributions across multiple headline categories, Dunn's test would help identify which specific headline categories were significantly different from each other.

\section{Limitations}
\subsection{Sampling Bias}\label{sec:sampling_bias}

Participants recruited through Prolific might not be a representative sample of the general population, leading to potential sampling bias. The results might be more reflective of individuals familiar with online tasks and surveys rather than a broader demographic. While we would therefore be hesitant to claim biases observed in our results are held by the general population, the participant's profiles likely align with those typically crowdsourced for identifying fake news in real-world applications. In particular, not all participants may exhibit the same level of conscientiousness as an experienced fact checker. Therefore, their inclusion in our results highlights the importance of algorithms that can handle such outliers.

Similarly, our recruitment of participants within the Anglosphere limits our ability to generalize to the whole world. We believe this to be mainly limiting when it comes to the observed biases, as it is likely that other populations hold other beliefs.

\subsection{Problem Selection}
Due to our desire to analyze the impact of biases and their mitigation, the headlines we selected focus on a subset of headlines which involve contrasting sensitive groups. Response patterns might differ when faced with broader sets of headlines wherein stereotypes do not play as strong of a role.

\subsection{Lack of Interaction}
The design choice to have participants operate in isolation, while intentional, means our study does not capture the potential influence of group dynamics, discussions, or peer pressure on decision-making. It should however be noted that commonly held stereotypes likely arise from group dynamics and discussions which occurred before participants completed our experiment. We leave to future work the exploration of explicit information exchange and how it should be managed.


\section{Preregistration Comparison}
We now compare and contrast this paper with the preregistered experimental design. While the objectives and experimental design we describe in this paper are in line with the preregistration, differences are essentially constrained to the language used to describe biases. The following compares and contrasts this paper with the preregistration in more detail.

\subsection{Research Questions}
Our preregistration outlined several primary and secondary research questions. Below, we detail how we addressed each question in this paper, referring to specific sections for in-depth analysis.\\

\noindent \textbf{does the within-factor algorithm have an effect on the accuracy for a subject group (i.e., is one algorithm better)?} We provide relevant results in \autoref{sec:alg_performance} and discuss them in section \textit{Collective Intelligence through Online Machine Learning}, . \\

\noindent \textbf{does the within-factor sensitive group have an effect on the advice or error of a subject individual (i.e., is an expert biased)?} We provide the relevant results in \autoref{sec:performance_by_sentiment}, and discuss them in \textit{Headline Categories and Implicit Biases}. Note that 'Sensitive group' and 'Headline Category' are used interchangeably to refer to the same feature.\\

\noindent In addition, we considered the following secondary questions:\\
\noindent \textbf{do people consistently misevaluate some headlines (either individual headlines or headlines pertaining to a particular protected group)? In other words, does the headline or sensitive group have an effect on the error and/or advice? These results can also be further evaluated in terms of participant demographics} This question is addressed through our results and discussions on the framing effect.\\

\noindent \textbf{do people tend to be more conservative/assertive in their answers? That is, are they more likely
to select the 'very' option? Again, results between demographic groups can be contrasted} In the main text, we discuss (a lack of) skepticism as a function of the headline category. In addition, we provide an answer to this question in the supplementary material. \\

\noindent \textbf{do demographically diverse groups enhance the wisdom of the crowd? I.e., is the performance of heterogeneous groups better than that of homogeneous groups.} We provide a discussion in the supplementary material. \\

\noindent \textbf{Is response time a predictor of advice quality or bias?} We provide a discussion in the supplementary material.

\section{Code availability}
Code necessary to reproduce our results is available at \url{https://github.com/axelabels/fakenews}.

\section{Data availability}
Data necessary for reproducing our results, including participant responses, is available at \url{https://doi.org/10.5281/zenodo.10794209}.

\section{Ethics considerations}

Due to the sensitive nature of the questions, participants were informed ahead of the questionnaire that ``Some of the questions presented in this study involve sensitive characteristics such as gender and ethnicity. Please be advised that exposure to scenarios involving these characteristics may potentially make you feel discomfort.". In addition, ethical approval was granted on February 16th 2023 by the host institution's ethics board before data collection. To ensure ethical compliance, informed consent was obtained from all participants. This consent process involved clearly explaining the purpose of the study, the types of data that would be collected, how the data would be used, and the participants' rights.

\section{Acknowledgements}
A.A. was supported by a FRIA grant (nr. 5200122F) by the National Fund for Scientific Research (F.N.R.S.) of Belgium. T.L. is supported by the F.N.R.S. project with grant number 40007793, the Service Public de Wallonie Recherche under grant n\textdegree 2010235–ARIAC by DigitalWallonia4.ai. A.A., T.L. and A.N. benefit from the support of the Flemish Government through the AI Research Program. T.L. and A.N. acknowledge the support by TAILOR, a project funded by EU Horizon 2020 research and innovation program under GA No 952215. E.F.D. is supported by an F.N.R.S Chargé de Recherche position, grant number 40005955.


\section{Competing Interests}
The authors declare no competing interests.

\bibliographystyle{unsrtnat}
\bibliography{sample}

\newpage
\FloatBarrier
\begin{appendices}

\section{Algorithms}\label{sec:algorithms}
The problem we study in this work can be formalized as a problem of bandits with expert advice \citep{auer2002nonstochastic}.

A straightforward approach in this setting is to perform static aggregation of participant opinions. In particular, decisions can be made by acting on an arithmetic mean of all opinions, resulting in a \textbf{majority vote}. Such votes are most effective when expertise is homogeneous and participants' errors are uncorrelated \citep{nicolas1785essai,Grofman1983}. Participants however often differ in their performance, and can show varying degrees of correlation due to shared experiences or stereotypes. Both factors degrade the performance of the simple average. By giving all group members an equal amount of influence, we dilute the knowledge of more accurate experts. The majority vote serves as a fundamental baseline for evaluating more sophisticated algorithms. Its simplicity and widespread use in decision-making scenarios make it a valuable point of comparison.
 
 If information about the experts' expected performance is available, votes can instead be weighted as a function of this performance estimate. The exponential weighting method \textbf{EXP4} \citep{auer2002nonstochastic} for example, maintains weights over the experts, which are updated to favor participants with higher performance. For each decision, it selects one of the participants according to its current probability distribution. This single participant's opinion is then acted upon. In this sense, the decisions made by EXP4 are not collective, as a single participant is decisive. 
 However, among algorithms that select a single participant, EXP4 has optimal theoretical guarantees and is straightforward to implement, leading to its inclusion in this study. 

The reliance on a single expert by EXP4 underscores a significant limitation: the lack of collective decision-making. Recent advancements have introduced algorithms like \textbf{MetaCMAB} \citep{abels2023dealing}, which, by employing a regression across all participant opinions, enables a more nuanced combination of expert advice. In doing so, MetaCMAB can optimally combine the opinions of diverse groups in order to enable better decisions. This approach allows MetaCMAB to outperform even the best individual participant, showcasing the potential of collective intelligence.

Nonetheless, MetaCMAB and similar algorithms assume uniform performance across different contexts (here, headline categories), an assumption rarely met in practice. Performance can vary significantly with the context, influenced by factors like prior experience and cognitive biases. Such heterogeneous performance permeates into collective decisions, meaning the optimal weighting of opinions also is a function of the headline category. To tackle such heterogeneity, the \textbf{ExpertiseTree} \cite{pmlr-v202-abels23a} learns to split the context space (consisting here of the headline categories) and then fits distinct models to each region. Note that regions are only split if it induces a better estimated performance. In theory, this allows ExpertiseTree to adapt the granularity of its models to the available expertise. 

\section{Framing Effect for EXP4}\label{sec:framing_exp}

\begin{figure}
\centering
\includegraphics[width=.4\textwidth]{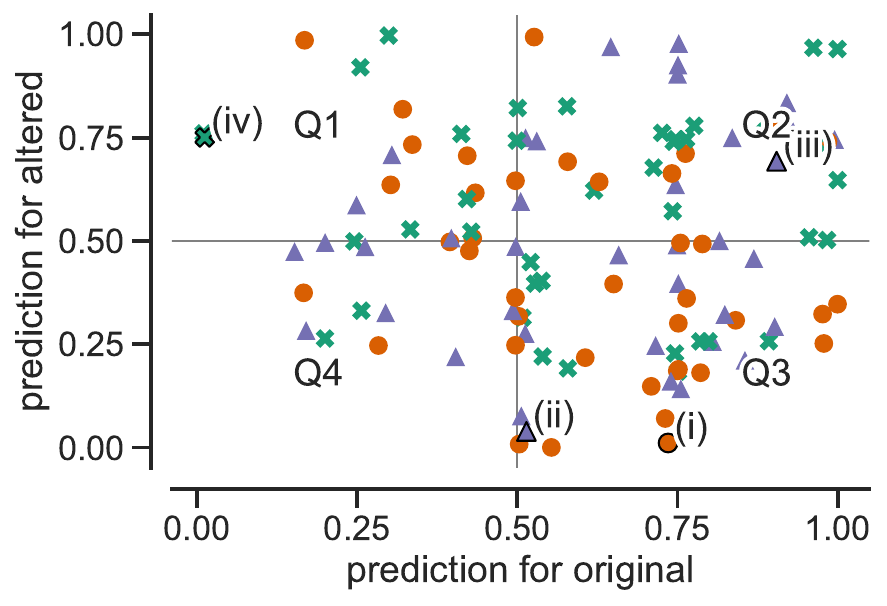}
\caption{Effects of framing in participants' responses for EXP4. This plot is analogous to those in \autoref{fig:scatter_real_altered}; the quadrants represent false stereotypes (Q1), positive framing effects (Q2), common knowledge (Q3), and negative framing effects (Q4).
Each point represents a distinct headline; the x-coordinate displays the expected response to its original form, and the y-coordinate shows the expected response to its altered form.}
\label{fig:scatter_exp4}
\end{figure}

We provide in \autoref{fig:scatter_exp4} the framing effects for EXP4 (analogously to those in \autoref{fig:scatter_real_altered}). EXP4 tends to concentrate its weights on a single group member, resulting in predictions which are often aligned with that group member's advice (and are thus close to $0, 0.25, 0.5,0.75, \text{or } 1$). In terms of shifts from the simple average, we find the following: Q1: $19\rightarrow 16$, Q2: $34 \rightarrow 41$, Q3: $52\rightarrow 49$, and Q4: $15 \rightarrow 14$. While there are slight shifts from Q3 to Q2, overall these framing effects are comparable to the average.

In terms of group biases, captured by the GEE summary in \autoref{tab:gee_exp4}, none of the features seem to have a significant impact, however the intercept indicates the model's error is generally large.

\section{Transposed Categories}\label{sec:experiment_2}
Our results in the main text suggest that headline categories induce different biases. For example, we found that age-related headlines gave rise to low skepticism, whereas headlines related to ethnicity gave rise to high skepticism. To eliminate the hypothesis that this is not due to the categories but instead is a result of the specific headlines presented in each category, we ran a secondary experiment wherein the sensitive groups were swapped. For example, we swapped a headline's age group by a gender group. If our hypothesis that the biased responses are due to the groups themselves, rather than to the specific headlines, is true, we would expect to find the same patterns of (lack of) skepticism in these altered headlines. In contrast, if the (lack of) skepticism was a result of the headlines themselves, we would find that when permuting categories, the categories susceptible to skepticism would also change. 

Our results suggest that our hypothesis holds, as we find that the number of age headlines in Q4 is low both when generating age headlines from gender headlines (2/80 headlines in Q4), and when generating age headlines from ethnicity headlines (3/80 headlines in Q4).

In contrast, ethnicity headlines are absent from Q2 when generated from age headlines (0/80 headlines in Q2) or from gender headlines (again 0/80 headlines in Q2). 

\section{GEE tables}
Tables \ref{tab:gee_experts} to \ref{tab:gee_expertisetree} summarize the GEE analyses in support of the group bias discussion of \autoref{sec:group_biases}. These tables analyze the effects of headline class (gender, ethnicity, age) and alteration status (genuine vs. altered) on error. Headline class variables indicate the headline's focus, with the baseline being age. ``Genuine" takes as value $1$ when the headline is unaltered. Interaction terms (e.g., Gender+Genuine) explore the combined effect of headline class and authenticity. 

Reference characteristics are age and altered. Therefore, for example, an estimate of $0.579$ for the intercept in \autoref{tab:gee_experts} indicates that for an altered headline conveying an outcome about age groups, the expected error is $0.579$. Other estimates are coefficients, such that, for example, a value of $-0.093$ for the ``Headline Class: Gender" parameter indicates that the expected error for altered headlines about gender is $0.579-0.093=0.486$. In addition, the associated $p$-value indicates that this difference is significant. We therefore conclude that, when using an average to aggregate, there is a significant difference in error between genuine and altered headlines. Conversely, for MetaCMAB for example (\autoref{tab:gee_metacmab}), the same parameter's estimate is $-0.036$, but with a $p$-value of $0.660$, suggesting this feature no longer has a significant effect. 

\begin{table*}[!htbp]
\centering
\begin{tabular}{lcccc}
\toprule 
Parameter & Estimate & Std. Error & t-value & $p$-value \\ 
 \midrule 
Intercept & 0.579 & 0.022 & 26.085 & $<$ 0.001 \\ 
Headline Class: Ethnicity & -0.135 & 0.035 & -3.866 & $<$ 0.001 \\ 
Headline Class: Gender & -0.093 & 0.047 & -1.972 & 0.049 \\ 
Genuine & -0.204 & 0.040 & -5.045 & $<$ 0.001 \\ 
Headline Class: Ethnicity:Genuine & 0.219 & 0.058 & 3.782 & $<$ 0.001 \\ 
Headline Class: Gender:Genuine & 0.155 & 0.063 & 2.435 & 0.015 \\ 
\bottomrule 
\end{tabular}
\caption{GEE summary for the average, predicting the aggregate's error as a function of headline characteristics.}
\label{tab:gee_experts}
\end{table*}

\begin{table*}[!htbp] 
\centering 
\begin{tabular}{lcccc} 
 \toprule 
Parameter & Estimate & Std. Error & t-value & $p$-value \\ 
 \midrule 
Intercept & 0.537 & 0.046 & 11.712 & $<$ 0.001 \\ 
Headline Class: Ethnicity & -0.088 & 0.057 & -1.553 & 0.120 \\ 
Headline Class: Gender & -0.036 & 0.081 & -0.440 & 0.660 \\ 
Genuine & -0.126 & 0.110 & -1.144 & 0.253 \\ 
Headline Class: Ethnicity:Genuine & 0.034 & 0.118 & 0.285 & 0.776 \\ 
Headline Class: Gender:Genuine & -0.084 & 0.170 & -0.492 & 0.622 \\ 
\bottomrule 
 \end{tabular} 
\caption{GEE summary for EXP4, predicting the aggregate's error as a function of headline characteristics.} 
\label{tab:gee_exp4} 
\end{table*}

\begin{table*}[!htbp] 
\centering 
\begin{tabular}{lcccc} 
 \toprule 
Parameter & Estimate & Std. Error & t-value & $p$-value \\ 
 \midrule 
Intercept & 0.431 & 0.027 & 15.819 & $<$ 0.001 \\ 
Headline Class: Ethnicity & -0.087 & 0.038 & -2.304 & 0.021 \\ 
Headline Class: Gender & -0.056 & 0.039 & -1.417 & 0.156 \\ 
Genuine & -0.079 & 0.035 & -2.274 & 0.023 \\ 
Headline Class: Ethnicity:Genuine & 0.134 & 0.023 & 5.780 & $<$ 0.001 \\ 
Headline Class: Gender:Genuine & 0.110 & 0.068 & 1.610 & 0.107 \\ 
\bottomrule 
 \end{tabular} 
\caption{GEE summary for MetaCMAB, predicting the aggregate's error as a function of headline characteristics.} 
\label{tab:gee_metacmab} 
\end{table*}

\begin{table*}[!htbp] 
\centering 
\begin{tabular}{lcccc} 
 \toprule 
Parameter & Estimate & Std. Error & t-value & $p$-value \\ 
 \midrule 
Intercept & 0.319 & 0.017 & 19.326 & $<$ 0.001 \\ 
Headline Class: Ethnicity & -0.053 & 0.033 & -1.606 & 0.108 \\ 
Headline Class: Gender & -0.043 & 0.037 & -1.154 & 0.249 \\ 
Genuine & 0.003 & 0.024 & 0.136 & 0.892 \\ 
Headline Class: Ethnicity:Genuine & -0.015 & 0.035 & -0.426 & 0.670 \\ 
Headline Class: Gender:Genuine & 0.029 & 0.026 & 1.080 & 0.280 \\ 
\bottomrule 
 \end{tabular} 
\caption{GEE summary for ExpertiseTree, predicting the aggregate's error as a function of headline characteristics.} 
\label{tab:gee_expertisetree} 
\end{table*}




\section{Secondary Questions}
\subsection{Inclination Towards Conservative or Assertive Responses: Contrast by Error and Demographic Group}

\begin{figure}
\centering
\begin{subfigure}[b]{0.45\textwidth}
\centering
\includegraphics[width=1\textwidth]{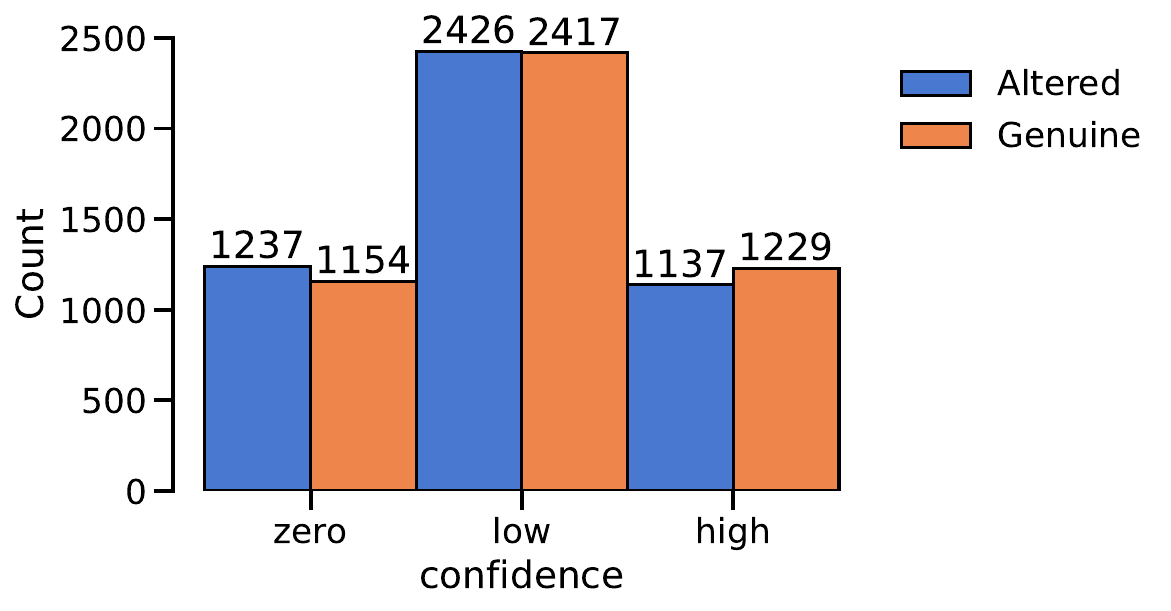}
\caption{Distribution of confidence levels as a function of altered status.}
\label{fig:confidence_histogram}
\end{subfigure}
\hfill
\begin{subfigure}[b]{0.45\textwidth}
\centering
\includegraphics[width=1\textwidth]{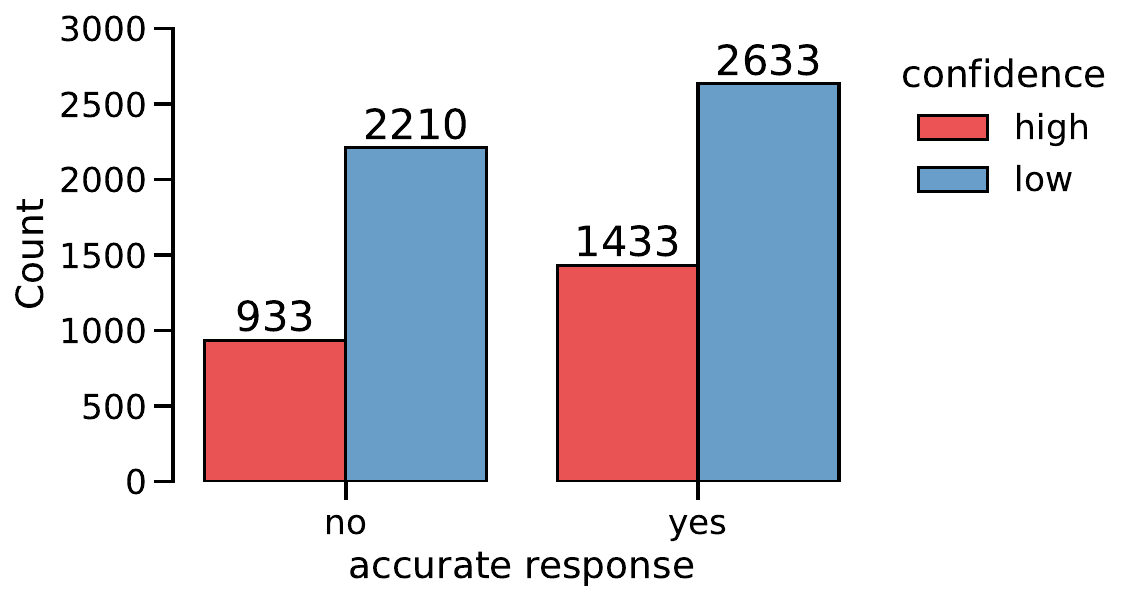}
\caption{Distribution of accuracy as a function of confidence level.}
\label{fig:accuracy_histogram}
\end{subfigure}
\caption{Confidence histograms}
\end{figure}

\begin{figure*}
\centering
\includegraphics[width=1\textwidth]{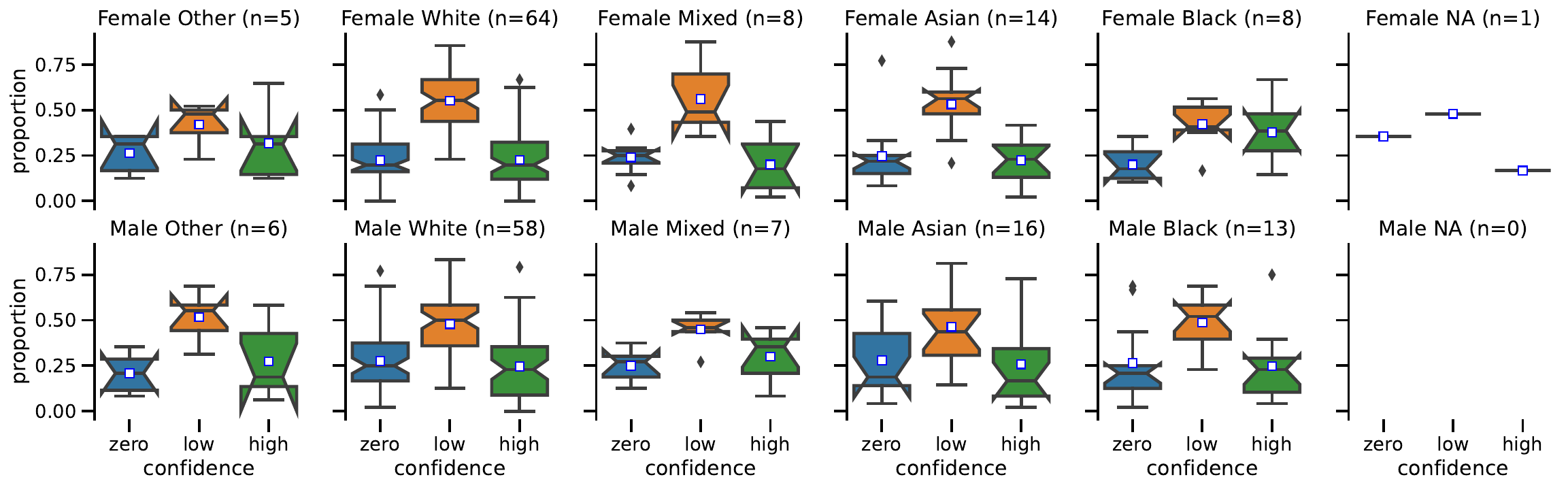}
\caption{Proportion of confidence categories displayed by participant demographic groups.}
\label{fig:confidence_boxplots}
\end{figure*}
People tend to be conservative in their answers, being more likely to answer (un)likely rather than very (un)likely (see Figure \ref{fig:confidence_histogram}).

Note that detailing by demographic groups (Figure \ref{fig:confidence_boxplots}) shows that in general, all demographic groups tend to more low confidence than high confidence. One exception to this observation is the group of black females which participated in this study, as they tend to express low and high confidence in equal measure. 
Furthermore, people do tend to answer more extremely when correct than when incorrect, indicating that their confidence estimates are somewhat calibrated. In particular, when maximally confident, participants have an expected accuracy of 60.6\%, while, when less confident, they have an expected accuracy of 54.2\%. 

While the difference is small, it suggests that weighing the participants' contribution by their own confidence is likely to be beneficial. Comparing the accuracy of a majority vote with a confidence-weighted majority vote, we find that the latter obtains an accuracy of $64.0\%$, while the former's accuracy is $54.8\%$. A Wilcoxon test suggests that this difference is significant ($p=0.0008$).

\subsection{Response Time as an Indicator of Accuracy}
\begin{figure}
\centering
\includegraphics[width=0.5\textwidth]{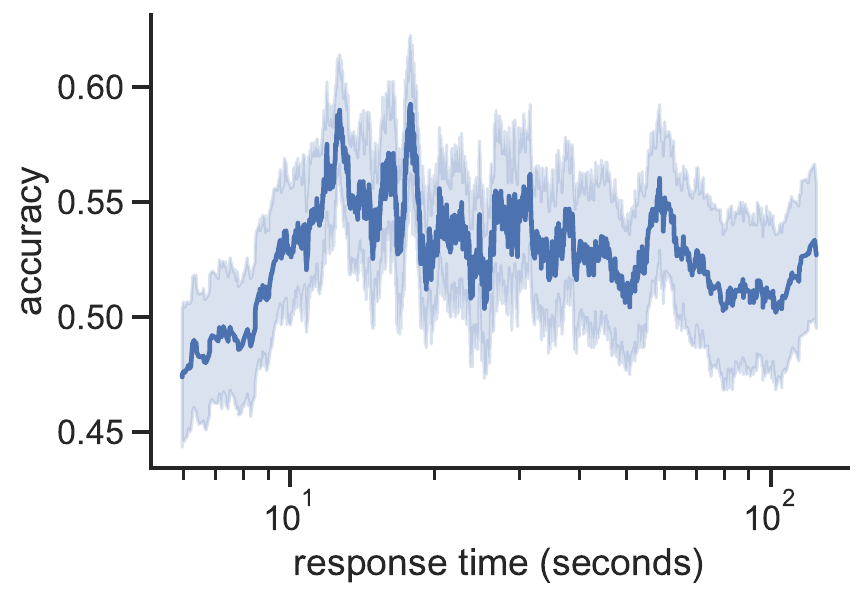}
\caption{Accuracy by response time. Responses are sorted by response time and then aggregated through a moving average with a window of size $100$. Shaded areas are the bootstrapped $95\%$ confidence intervals.}
\label{reward_by_time}
\end{figure}

We observe two trends in Figure \ref{reward_by_time}. First, the fastest response times induce the lowest accuracy. This suggests that for those response times, users simply provided answers without taking the prompts into account. Conversely, as response time increases, the error rate also tends to increase. These slower response time could be a sign of greater uncertainty, which correlates with more mistakes. Finally, the rightmost points have a slightly higher performance, however, given the large response times involved (over $5$ times the standard deviation), these responses were likely given after participants switched off the questionnaire. 

\subsection{Performance of Demographically Diverse Groups}

While we found that women perform slightly better than men, this individual improvement is not reflected in an improved performance of groups; we find that acting on the average of all female responses is not superior to acting on the average of all male responses (based on a Wilcoxon test). This is likely because, while women tend to perform better on average, they are more strongly correlated than men; men have an average Pearson correlation coefficient of $0.178$, while women have an average correlation coefficient of $0.220$. In practical terms, higher correlation within groups will result in group members making more of the same errors, which are then reinforced by an average aggregate. In contrast, lower correlation means individuals tend to make errors on different instances, which are more easily subdued by an average. 

We might therefore wonder whether mixing demographic groups is a better approach to maximizing group performance than selecting the group with the best individuals. Unfortunately we found no significant improvement from mixed groups. While we might hope that mixed groups would be less correlated than the individual groups which compose them, we found that instead the correlation of a mixed group was approximately equal to the average correlation of the individual groups. In particular this means that a mixed group is typically more correlated than the least correlated individual group. For example, the correlation of a group with equal number of women and men is $0.194$, approximately halfway the correlation values of the individual groups given above.

\end{appendices}

\end{document}